%% file: 03_ARXIV.tex
\documentclass[a4paper,12pt]{scrartcl}
\usepackage[utf8]{inputenc}
\usepackage{mathtools}
\usepackage{amsmath, amsfonts, amssymb, bm, dsfont}
\usepackage{bbm}
\usepackage{xcolor}
\usepackage{graphicx}
\usepackage[left=2.5cm,right=2.5cm,top=2.5cm,bottom=2.5cm]{geometry}
\usepackage[ruled,vlined]{algorithm2e}
\usepackage{booktabs}
\usepackage{comment}
\date{}
\usepackage{placeins}
\usepackage{tikz}
\usetikzlibrary{shapes,arrows,chains}
\usepackage{natbib}
\usepackage{amsthm} % for unnumbered remark
\newtheorem{proposition}{Proposition}[section]

\newtheorem*{remark}{Remark}
\usepackage{booktabs}
\usepackage[colorlinks = true,
            linkcolor = teal,
            urlcolor  = teal,
            citecolor = teal,
            anchorcolor = teal]{hyperref}

\newcommand{\btheta}{\boldsymbol{\theta}}
\newcommand{\bomega}{\boldsymbol{\omega}}
\newcommand{\bpi}{\boldsymbol{\pi}}

\newcommand{\bv}{\boldsymbol{v}}

\newcommand{\prob}[1]{\mathbb{P}\left[#1\right]}

\newcommand{\simiid}{\overset{iid}{\sim}}
\newcommand{\simind}{\overset{ind}{\sim}}
\newcommand{\Beta}{\mathrm{Beta}}

\newcommand{\E}[1]{\mathbb{E}\left[#1\right]}

\newcommand{\Dirichlet}{\mathrm{Dirichlet}}
\usepackage{url} % not crucial - just used below for the URL 

\title{Multiomics Tissue Segmentation via Spatially-Informed Nested Biclustering Methods}
\date{}
\author{Francesco Denti\footnote{Department of Statistical Sciences, University of Padua} , 
    Cecilia Balocchi\footnote{School of Mathematics, University of Edinburgh} ,\\
    Vanna Denti\footnote{Proteomics and Metabolomics Unit, Department of Medicine and Surgery, University of Milano-Bicocca} , and 
    Giulia Capitoli\footnote{Bicocca Bioinformatics Biostatistics and Bioimaging 
    B4 Center, Department of Medicine and Surgery, University of Milano-Bicocca}{ }\footnote{Biostatistics and Clinical Epidemiology, Fondazione IRCCS San Gerardo Dei Tintori}}

\begin{document}

  \maketitle

\vspace{-1.5cm}
\begin{abstract}
	Matrix-Assisted Laser Desorption/Ionisation Mass Spectrometry Imaging (MSI) is a powerful technique for spatially resolved molecular profiling and cancer biomarker discovery. Recent advances, including a novel multiomics workflow, enable multiple rounds of MSI on the same tissue section, extracting diverse molecular classes, e.g., lipids, peptides, and N-glycans, while preserving spatial resolution. This innovation is particularly valuable for studies with limited tissue sample availability, such as rare diseases or small tumors. However, the resulting data are high-dimensional, spatially structured, and the various molecular types share the same pixel grid. To address these challenges, we propose Poseidon, a Bayesian nonparametric nested biclustering model that simultaneously segments the common tissue pixels and clusters molecular signals within each molecular class, leveraging the shared spatial structure. A separately exchangeable framework is first considered, and then extended to handle spatial data via hidden Markov random fields.
	For scalability, we implement an efficient mean-field variational inference algorithm tailored for multi-dataset analysis. After validating the efficacy of our method on simulated scenarios, we demonstrate the applicability of our model in a real-world case study, where multiomics measurements were performed on kidney tissue affected by clear cell renal cell carcinoma. The nested, hierarchical structure of Poseidon, combined with its principled inferential framework, allows the extraction of interesting biological insights, such as clear tissue segmentation and biomarker detection. 
\end{abstract}

%  Please place your key words in alphabetical order, separated
%  by semicolons, with the first letter of the first word capitalized,
%  and a period at the end of the list.
%

\textit{Keywords}: Clear cell renal cell carcinoma; Dirichlet process; Potts model; Shared atoms nested models; Variational inference.

\include{01_MAIN_skeleton}

\bibliographystyle{plainnat}
\bibliography{sepexch}

\clearpage
\begin{center}
\Huge
    \textbf{Supplementary Material}
\end{center}

\include{02_SUPP_skeleton}

\end{document}

%% file: 01_MAIN_skeleton.tex
\section{Introduction}
\label{sec:intro}

Over the last decade, \emph{Matrix-Assisted Laser Desorption/Ionisation} (MALDI) Mass Spectrometry Imaging (MSI) has emerged as a key technology for in-situ molecular profiling and the discovery of cancer biomarkers. This technique is devised to extract \emph{mass spectra}, i.e., the molecular abundances of a specific molecular class (e.g., lipids) in different locations of a biological sample \citep{rohner2005maldi}. 
Broadly speaking, the MALDI-MSI extraction works as follows. First, a slide of a biological sample is covered with a chemical compound called \emph{matrix}, which separates a specific class of molecules from the rest and makes it ready to be detected. Then, the slide is divided into a grid of virtual pixels, typically 10–50 $\mu m$ in size, each of which is hit with a laser beam. The instrument separates different molecules - the \emph{analytes} of interest -  depending on their mass-to-charge index, called \emph{m/z}, and acquires a mass spectrum that collects the abundance (in terms of intensities) for each \emph{m/z}. 

\begin{figure}[t]
	\centering
	\includegraphics[width=\linewidth]{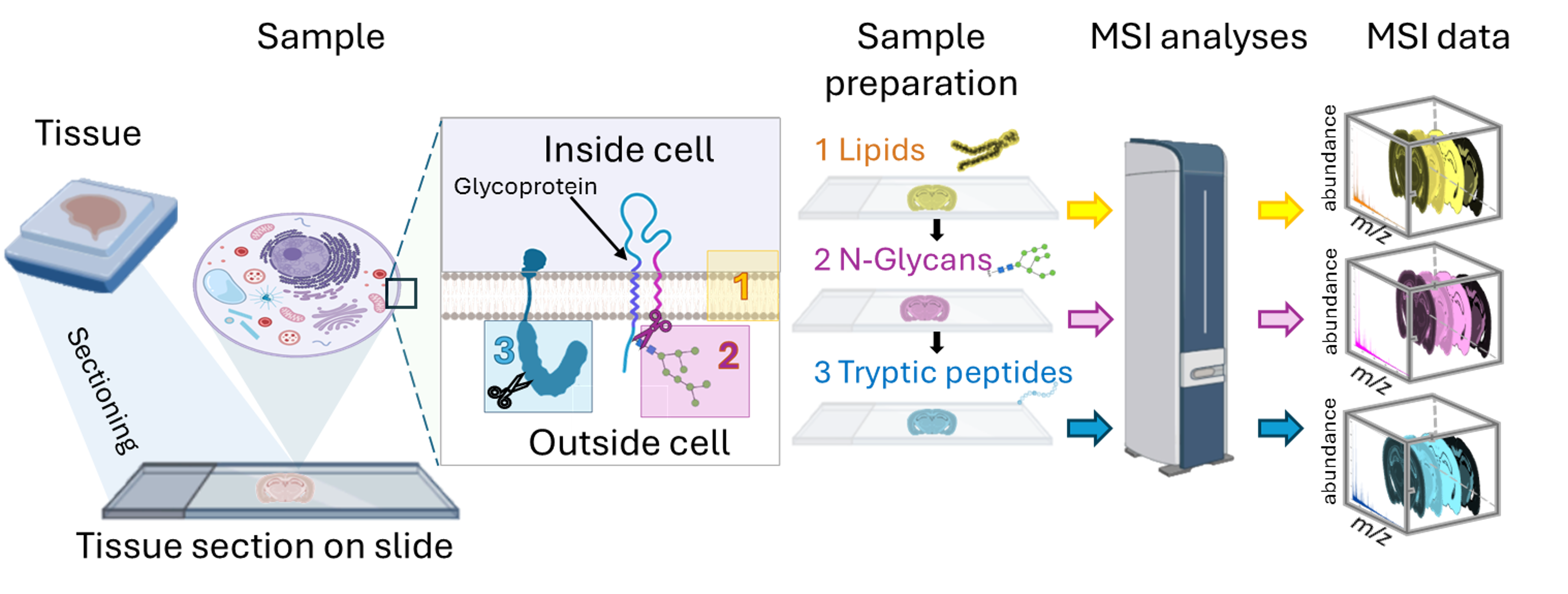}
	\caption{A depiction of the \emph{multiomics workflow} developed in \cite{denti2022spatial}. The \emph{same} tissue undergoes multiple consecutive MSI analyses with MALDI, yielding different MSI datasets -- one for each molecular class.}
	\label{fig:fig1_multiomics1}
\end{figure}

Recent advances in MSI have improved both the precision of molecular detection and the reproducibility of experiments. This has expanded the field of spatial proteomics and enabled a finer characterization of tumor microenvironments \citep{krestensen2023state, moore2023prospective, zhang2024mass}. Of particular significance is a recent innovation by \citet{denti2022spatial}, who introduced a novel \emph{multiomics workflow} allowing multiple rounds of MSI acquisition on the \emph{same} tissue section. By sequentially applying and removing different \emph{matrices}, this method enables the extraction of analyte intensities from distinct molecular classes (such as, e.g., lipids, peptides, and N-glycans),
while minimizing tissue degradation; as a result, each molecular class yields a distinct MSI dataset acquired over \emph{a common spatial grid}. Figure~\ref{fig:fig1_multiomics1} illustrates this pipeline. This integrated approach is especially valuable in contexts with limited clinical samples, such as rare diseases or small tumors, where obtaining comprehensive molecular information from a single section maximizes biological insight.
However, these experimental advances introduce new analytical challenges. 
Multiomics MSI data are inherently high-dimensional and spatially structured, and the novel \emph{multiomics workflow} further increases complexity by generating nested data structures. The pipeline outputs multiple MSI matrices (one for each molecular class), which share the same spatial grid of pixels (columns), but differ in the analytes they measure (rows). 
A core objective in this context is tissue segmentation, which involves identifying biologically meaningful subregions of a sample based on their spectral characteristics, to uncover subtle molecular variations that may be missed by morphological inspection alone.

Current approaches typically rely on linear and nonlinear dimensionality reduction techniques to extract simplified representations of MSI data \citep[see][]{zhvansky2021comparison}. Clustering methods such as k-means and bisecting k-means remain common in the field. 
More sophisticated segmentation strategies have also been proposed over the past decade, aiming to incorporate spatial structure and domain-specific features. As notable examples, see \cite{Alexandrov2011,guo2019unsupervised}, the latter employing model-based Gaussian Mixture Models, which have shown promising performance for MSI data \citep{prasad2022evaluation}. 
See also \citet{Verbeeck2019, Tuck2022} for a comprehensive review. However, these methods are generally designed for single-matrix inputs and do not account for the shared spatial structure, nested organization, or molecular heterogeneity across analyte types.

In this paper, we propose Poseidon (\textit{Potts Over Separate Exchangeability for the Integration of Different Omics Nested data}), a novel Bayesian nonparametric (BNP) method for the analysis of multiomics MSI data based on \emph{nested biclustering}, which simultaneously segments spatial pixels and clusters molecular signals within each tissue subregion. Our approach integrates the multiple MSI datasets while explicitly accounting for the shared spatial grid and underlying dependencies across pixels. In particular, Poseidon builds on the concept of separate exchangeability~\citep[see][and the references therein]{Lin2021}, providing a principled foundation for modeling complex nested structures in high-dimensional omics data. This framework naturally induces a two-layer partitioning structure, similar to that discussed in~\citet{Lee2013}. 
%(see also~\citet{Moran2021} for related biclustering approaches).
Our proposed method contributes in three key ways. First, (i) by jointly clustering pixels and mass-to-charge (\emph{m/z}) values, the model identifies signal groups with elevated or suppressed activation, effectively distinguishing biologically meaningful peaks from background noise and enabling the extraction of latent molecular features. This allows the extraction of latent feature representations that support the spatial segmentation of the tissue.
Second, (ii) it accommodates the nested structure of multiomics MSI data by integrating multiple molecular classes into a unified framework, allowing comparisons across analyte types measured on the same tissue. Finally, (iii) to ensure scalability and efficient inference in high-dimensional settings, we implement a tailored variational approximation to the posterior distribution, drawing on recent advances in mean-field variational inference for hierarchical, nested Bayesian models~\citep{Blei2017, Dangelo2023}.

\subsection{Clear Cell Renal Cell Carcinoma MALDI-MSI Data}
\label{subsec:data}

\begin{figure}[t]
    \centering
    \includegraphics[width=1\linewidth]{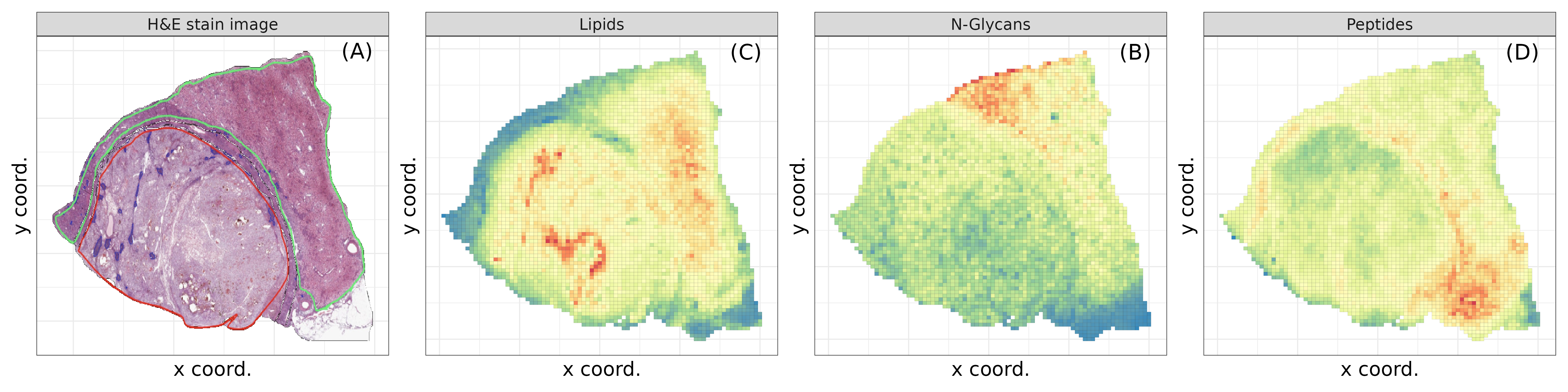}
    \includegraphics[width=1\linewidth]{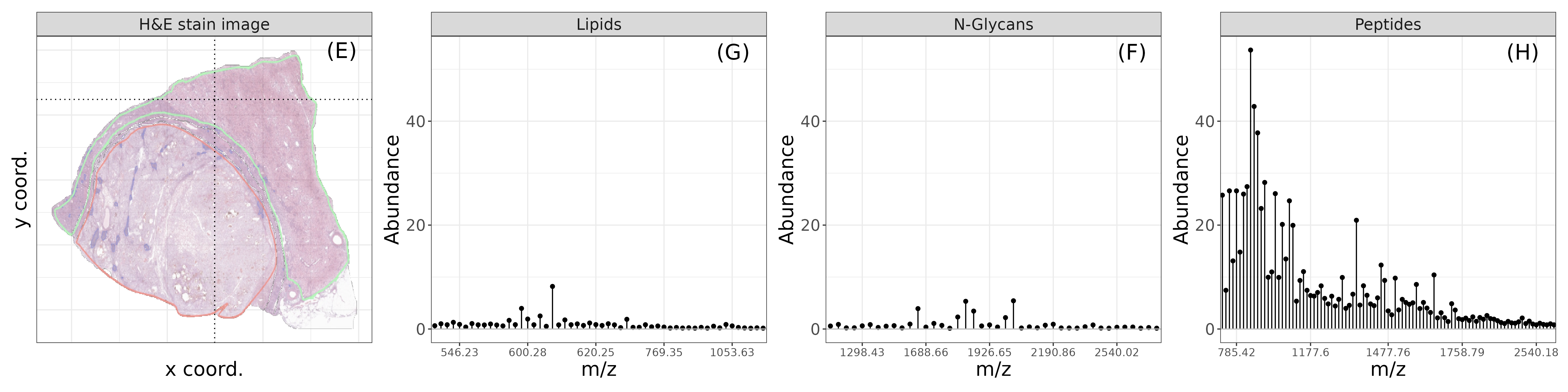}
    \caption{The top row displays the Hematoxylin and Eosin stained image of the ccRCC tissue used in the analysis (A) and heatmaps displaying the (normalized) median abundance across the tissue sample for lipids, N-glycans, and peptides (B-D). 
    The bottom row reports, for a specific highlighted pixel (E), abundances of the signals for the different molecular classes (F-H).}
    \label{fig:fig2_data}
\end{figure}

Our application focuses on MALDI-MSI measurements extracted from a human tissue sample diagnosed with clear cell renal cell carcinoma (ccRCC), shown in panel (A) of Figure~\ref{fig:fig2_data}. The tissue is annotated as follows: the green region indicates the \textit{healthy cortex}, the blue outlines the \textit{capsule}, and the red highlights the \textit{tumoral nodule}. A more detailed segmentation is provided in Figure~\ref{suppfig:S1_ccrcc_legend} of the Supplementary Material. Renal cell carcinoma is among the ten most frequently diagnosed cancers globally, with the clear-cell subtype accounting for roughly 75\% of cases and the majority of related mortality \citep{hsieh2017renal}. The prognosis is often poor due to high recurrence rates and resistance to conventional chemotherapy and radiotherapy. 
Investigating the phenotypic characteristics of the cells within these tumor environments could uncover novel molecular pathways, suggesting alternative treatment strategies \citep{hsieh2018genomic}.

Using the multiomics pipeline described in \citet{denti2022spatial}, mass spectra were acquired for three distinct molecular classes on the same renal tissue section: \textit{lipids, N-glycans}, and \emph{tryptic peptides}. These classes were selected for their complementary biological roles. Indeed, lipids are the major constituents of cell membranes and serve as a major energy source supporting cell growth. They also function as molecular messengers and may contribute to drug resistance \citep{bian2021lipid, Fu2021, Shafiee2020}. {N-glycans} are among the most common modifications of proteins, primarily involved in directing proteins to specific cellular compartments~\citep{stanley2022n}. Finally, it is known that proteins play diverse roles, including regulating cell growth, controlling tissue architecture, and guiding cellular replication \citep{gobena2024proteomics}. In this protocol, proteins are enzymatically cleaved by trypsin to produce tryptic peptides, enabling more detailed proteomic analysis. Together, these three molecular classes provide a multi-layered perspective on tumor biology and may enhance biomarker discovery when analyzed in an integrated spatial framework, as shown in previous spatial multiomic datasets~\citep{balluff2019integrative}.

In the preparatory phase, the raw spectra extracted with MALDI underwent routine pre-processing steps to account for physiological, instrumental, and analytical variability. Specifically, the spectra followed the biological five-step protocol described in~\citet{Capitoli2024}, summarized in Section~\ref{supp:sec_rawdata} of the Supplementary Material. These pre-processed signals were then stored in three separate matrices, corresponding to lipids, N-glycans, and peptides. In each matrix, columns represent spatial pixels (shared across matrices). At the same time, rows correspond to analyte-specific $m/z$ signals (which differ by molecular class). Each matrix contains approximately 100 rows and over 4,500 columns.
Figure~\ref{fig:fig2_data} provides a visual overview of the data. Panels (B), (C), and (D) show the normalized median abundance of each molecular class across the tissue section. These images highlight molecular heterogeneity that is not always apparent from histology alone. Panels (F), (G), and (H) illustrate spectral profiles for a single pixel (highlighted in panel E), showing the abundance levels across $m/z$ indices for each molecular class. Note that normalization was performed within each molecular class, so intensities are not directly comparable across the three panels.

By applying our proposed method to these data, we aim to identify molecular signatures and spatial subregions within the renal tissue that distinguish tumor from non-tumor areas, offering insights into the heterogeneity of clear cell renal cell carcinoma. Moreover, the biclustering approach aligns with our broader goal of reducing data complexity and delivering interpretable, low-dimensional summaries to support pathologists in downstream analyses.

The rest of the paper is organized as follows. In Section~\ref{sec:method}, we introduce our biclustering model, first outlining the core method for a single molecular layer and then extending it to jointly handle multiple analytes. Section~\ref{sec:VB} presents the variational inference 
ithm used for posterior estimation. In Section~\ref{sec:simulation}, we evaluate the performance of our method through simulation studies. Finally, in Section~\ref{sec:application}, we apply the model to the ccRCC dataset and analyze the resulting spatial and molecular patterns. Section~\ref{sec:conclusions} concludes.

\section{A finite-infinite separately exchangeable model for MALDI-MSI data} \label{sec:method}

\begin{figure}[t]
    \centering
    \includegraphics[width=\linewidth]{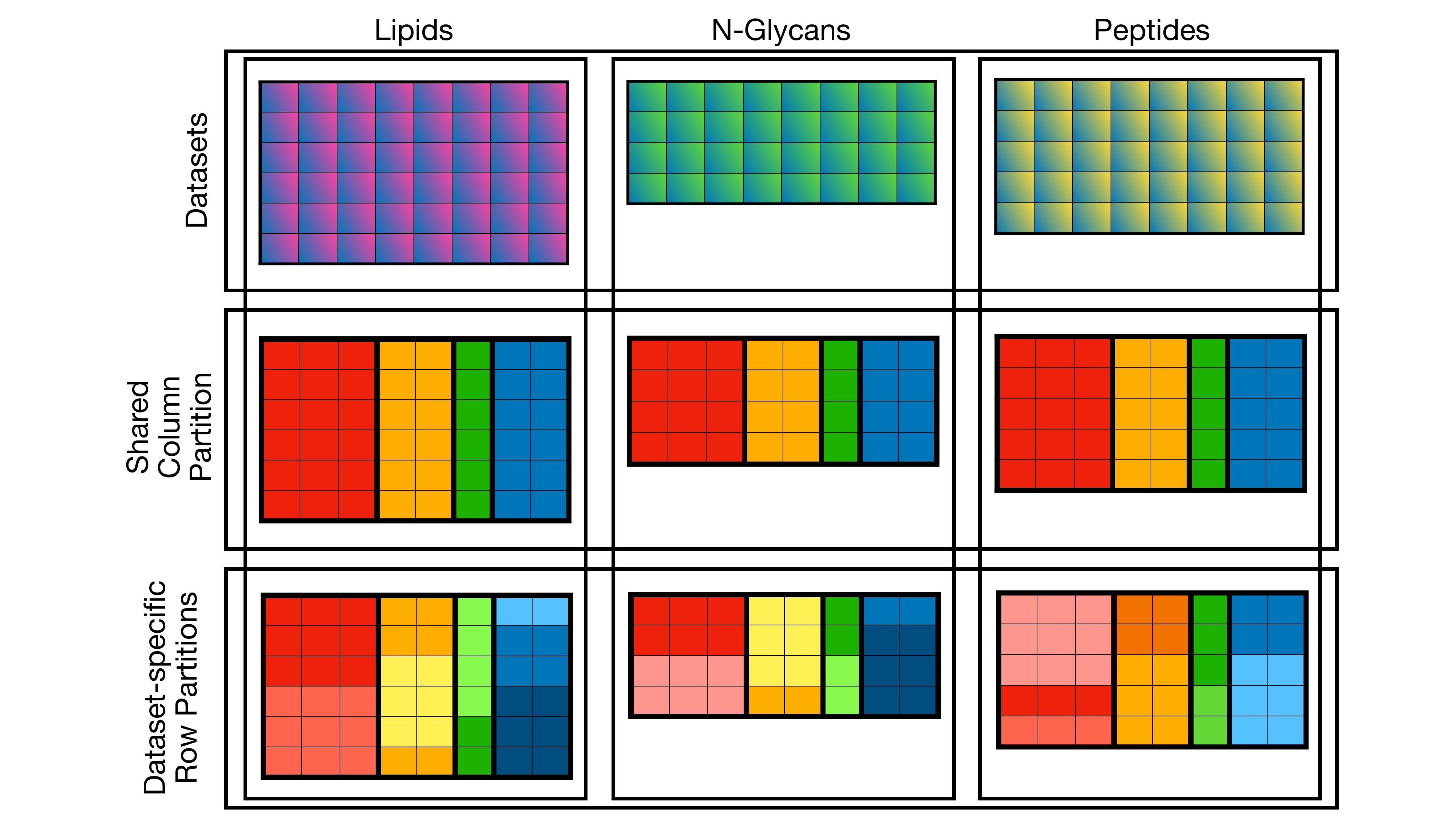}
    \caption{A diagram summarizing the partition structure we aim to recover. The partition over the columns is shared across datasets. The partition over the rows is dataset-specific.}
    \label{fig:fig3_clustScheme}
\end{figure}

In line with the data structure outlined in Section~\ref{subsec:data}, we consider a collection of $T$ datasets $\boldsymbol{\mathcal{Y}}=\{\boldsymbol{Y}^{(t)}\}_{t=1}^T$, one for each analyzed molecular level. Then, $\boldsymbol{Y}^{(t)}$ denotes the MALDI-MSI abundance matrix of the $t$-th molecular class, which is characterized by $N^{(t)}$ rows, representing the specific analytes, and $J$ columns, representing the pixels. We remark that a crucial aspect of the data is that the number of columns is the same across all the datasets.
Our main aim is to estimate a nested random partition to perform image segmentation, i.e., inducing a \emph{shared partition} of the $J$ pixels driven by the distributional characteristics of the molecular abundance within each location across datasets and, at the same time, inducing an idiosyncratic, \emph{dataset-specific partition} of the signals, different in each datasets. Figure~\ref{fig:fig3_clustScheme} summarizes the objective of our model. In what follows, we begin by discussing our model for the single-dataset case ($T=1$) and then extend it to the multiple-dataset scenario.

\subsection{The single-dataset case: Pose}
Let us focus on a single dataset, say $\boldsymbol{Y}$, with $N$ rows and $J$ columns. 
From a modeling perspective, we see our data as organized into distinct but related groups defined by the pixels, or columns. This setting is typical of nested designs, an area of research that has thrived in recent years within the Bayesian literature \citep[see, for example,][]{Rodriguez2008, Camerlenghi2019, Denti2021, Balocchi2022, Beraha2021, lijoi2023}. 
A common feature of nested models is that they aim at simultaneously clustering the groups and the observations; for matrix data where columns define groups, this translates into simultaneously clustering the columns and the matrix entries.
Most of these nested models were proposed under the hypothesis of \textit{partial exchangeability}, where data are assumed to be exchangeable within the different subpopulations. Consequently, the joint probability distribution of observations remains unchanged under any permutation of data points within the same subpopulation. As a consequence, under partially exchangeable nested models, the probability of clustering matrix entries does not depend on their row identity, and these models completely disregard the information conveyed by the specificity of each row, as \cite{Lin2021} have recently underlined. 
Indeed, this feature is problematic for us, as each row of our matrix has the same meaning (i.e., it represents the same $m/z$ value) across all the columns. 
Using a partially exchangeable model in our framework would mean clustering the pixels that have similar abundance distribution by simply considering the proportion of \emph{m/z} that have high vs low activation levels, regardless of \textit{which} \emph{m/z} achieve these activation levels. 
This might lead to an undesirable situation where, for example, two pixels are clustered together because they both have 5\% of the signals that are highly activated, 
even though the high activation is achieved by different sets of \emph{m/z} in the two pixels. Such a segmentation would lead to an important loss of information.

\textbf{Separate exchangeability.} To overcome this issue, similarly to \cite{Lin2021}, we start by considering a nested model in a \emph{separately exchangeable} setting, so that the identity of each row is preserved across the groups. We will introduce a set of column-specific clusters and, for each of them, a set of row-specific clusters, which are named, respectively, column clusters (CCs) and row clusters (RCs). Specifically, for each column $j = 1, \ldots, J$, let $C_j$ denote the CC membership label, with $C_j \in \{1,\ldots, K\}$.
Given the vector of CC allocations $\boldsymbol{C}=\left(C_1,\ldots, C_J\right)$, separate exchangeability implies that for each CC $k$, all elements in a row are restricted to belong to the same RC. In other words, for each row $i = 1, \ldots, N$ and \emph{every possible CC} $k$, we consider a RC membership label $R_{i,k}\in \{1, \ldots, L\}$ that is common to all entries $(i,j)$ whose columns are in the same CC characterized by $C_j = k$. Such RC allocation is considered for all possible CC realizations, with $\boldsymbol{R}_k=\left(R_{1,k},\ldots, R_{N,k}\right)$ for all $k$.

Let $f(y|\theta)$ be a suitable parametric density for the data, and $\{\theta^*_l\}_{l=1}^L$ denote a set of RC-specific parameters, common across all CCs (\emph{common atoms}). We define the distribution of a matrix entry $y_{i,j}$ given its CC assignment $C_j = k$ and RC assignment $R_{i,k} = l$, as  $y_{i,j} \vert \theta^*_l \sim f(y_{i,j}\vert \theta^*_l)$. 
The RC assignment variables $R_{i,k}$ are categorical random variables with CC specific parameters, such that $\mathbb{P} \left[ R_{i,k}=l \vert \boldsymbol{\omega}_k \right] = \omega_{l,k}$, where $\boldsymbol{\omega}_k = (\omega_{1,k}, \ldots,\omega_{L,k})$ is a vector of probabilities, for all $k$. 
In the most basic model specification, which will be revised in the next section, we also consider the CC assignment to be independent categorical random variables with probabilities $\boldsymbol{\pi}$, i.e., $\mathbb{P} \left[C_j=k \vert \boldsymbol{\pi} \right] = \pi_{k}$. 
Finally, we assume that $\theta^*_l \overset{i.i.d.}{\sim} H$, for $l = 1,\ldots,L$, where $H$ is a diffuse base measure.
Thus, we can express our model in terms of the cluster assignment variables $R_{i,k}$ and $C_j$, given the vectors of categorical probabilities:
\begin{equation}
\begin{gathered}\label{eq:se_clusterassignment}
    \theta^*_{l}  \overset{i.i.d.}{\sim} H, \\
    C_j \vert \boldsymbol{\pi} \sim \textrm{Cat}_K(\boldsymbol{\pi}), \quad \quad
    R_{i,k} \vert \boldsymbol{\omega}_k \sim \textrm{Cat}_L(\boldsymbol{\omega}_k),\\
    y_{i,j} \vert C_j, R_{i,C_j}, \{ \theta^*_l \}_l \overset{ind.}{\sim} f( y_{i,j} \vert \theta^*_{R_{i,C_j}} )
    % y_{i,j} \vert C_j = k, R_{i,k} = l, \theta^*_l &\sim f( y_{i,j} \vert \theta^*_{l} )
\end{gathered}
\end{equation}
Note that the matrix entries $y_{i,j}$ associated to the same RC $l$, i.e., the observations belonging to the set $\{ y_{i,j}: R_{i,C_j} = l \}$, are assumed to be exchangeable; in other words, they are modeled as conditionally i.i.d. from $f(\cdot \vert \theta^*_l)$.

\begin{remark}    
Drawing a parallel with the partially exchangeable nested model, we could describe this model by introducing column-specific random measures $G_j$, assumed to be distributed according to a discrete random measure $Q = \sum_{k\geq1} \pi_k \delta_{G^*_k}$. The discreteness of $Q$ induces an implicit clustering on the columns, and we could define $C_j = k$ when $G_j = G^*_k$.
For every possible CC $k$, we could consider row-specific parameters $\theta_{i,k}$, drawn i.i.d. from $G^*_k$. When the $G^*_k$'s are also discrete, with $G^*_k = \sum_{l\geq 1} \omega_{l,k}\delta_{\theta^*_{l}}$, a partition is also induced on the $\theta_{i,k}$ parameters, leading to the definition of the RC assignments $R_{i,k}$.
\end{remark}

Essentially, we cluster together different \emph{m/z} if they have similar activation levels (a trait captured by the kernel parameters $\{\theta^*_l\}_{l=1}^L$), and we group together different pixels if they have similar activation profiles, as represented by the vector of cluster assignment of the \emph{m/z}.
Note that according to the specification in~\eqref{eq:se_clusterassignment}, the distributional law of the data matrix $\boldsymbol{Y}$ is invariant to permutations of entire rows and columns. Moreover, given the nested nature of the model, the clustering of signals is affected by the CCs, as the cluster assignment of a given signal remains the same for all pixels belonging to the same CC.

Which prior distributions are selected for $\boldsymbol{\omega}_k$ (the \emph{observational weights}) and $\boldsymbol{\pi}$  (the \emph{distributional weights}) represents a crucial modeling choice. In \cite{Lin2021}, the authors consider a fully nonparametric approach, allowing both $L$ and $K$ to be infinite, and adopt stick-breaking (SB) Dirichlet process (DP) representations for both weights, inducing a \textit{common atom} structure. Here, we consider a finite-infinite version of the model, resulting in a \textit{shared atom} model, following \cite{Dangelo2023}, and 
we restrict $L$ to be finite, while $K$ is allowed to be infinite.
Specifically, we adopt sparse, symmetric Dirichlet distributions with parameters $\boldsymbol{b}_0=b_0\mathds{1}_L$ for the row mixture weights. Thus, for all $k$, $\boldsymbol{\omega}_k \sim Dir_L(\boldsymbol{b}_0)$. 
In the following, we will show how this simple change has a serious impact on the prior probability of co-clustering rows. Similarly to \cite{Lin2021}, we assume that $\boldsymbol{\pi}$ is drawn from a SB process with concentration parameter $\alpha$, denoted with $\textrm{GEM}(\alpha)$, which is equivalent to setting $\pi_1=v_1$ and $\pi_k = v_k\prod_{g<k}(1-v_g)$ where $v_j \overset{i.i.d.}{\sim} \textrm{Beta}(1,\alpha)$ for all $j \geq 1$.

\textbf{Pose: Potts over Separate Exchangeability}. The assumption of prior independence in the CC membership labels $\boldsymbol{C}_j$, which would be implied under separate exchangeability, could be naive in the context of spatial data. In fact, the columns in our dataset represent pixels of an image, which are likely to be spatially correlated. Therefore, we need to elicit a distribution that allows clustering of the columns, while conveying their spatial information.
Specifically, we assume that the clustering assignment for the generic pixel $j$ is influenced by the cluster memberships of its adjacent pixels. 
This structure can be recovered using Markov Random Fields \cite[MRF,][]{Julian1986}. 
Here, we adopt the BNP-MRF prior introduced by \cite{Lu2020}, which extends the classic BNP models for clustering using a Markov random field. In detail, this distribution assumes that
$p(C_j \mid \boldsymbol{C}_{-j},\boldsymbol{\pi},\beta) \propto \pi_{C_j} \cdot \exp(\beta \sum_{q \in \partial_j} \mathds{1}_{\{C_q=C_j\}} \,)$, 
where $\boldsymbol{\pi} \sim \textrm{GEM}(\alpha)$, and $\partial_j$ denotes the set of neighbors of unit $j$, i.e., the set of pixels adjacent to pixel $j$. The quantity $\beta$, known as the inverse temperature, governs the level of influence that the neighboring cluster memberships have on a pixel, and can be treated as a fixed hyperparameter or modeled as a random quantity. This conditional distribution 
can be equivalently stated as a joint distribution for the whole vector of cluster memberships $\boldsymbol{C}$, as
 \begin{equation}
    p(\boldsymbol{C}\mid\boldsymbol{\pi},\beta) = \frac{1}{\mathcal{K}_{\boldsymbol{C}}(\boldsymbol{\pi},\beta)}  \left( \prod_{j=1}^J \pi_{C_j} \right)  \cdot \exp\left[ \beta \sum_{j=1}^J\sum_{q\in \partial_j}\mathds{1}_{\{C_q=C_j\}} \right] 
    \label{eq:potts}
 \end{equation}
where $\mathcal{K}_{\boldsymbol{C}}(\boldsymbol{\pi},\beta)$ is an intractable normalizing constant. Therefore, we redefine the prior distribution for the CC membership $\boldsymbol{C}$ as \eqref{eq:potts}, and denote it with $\boldsymbol{C} \vert \boldsymbol{\pi},\beta \sim \textrm{BNP-MRF}(\bpi,\beta)$.

We refer to the model in~\eqref{eq:se_clusterassignment} with the CC prior redefined as in~\eqref{eq:potts} as the \emph{Potts Over Separable Exchangeability} (Pose) model. Note that under Pose, the model is no longer separately exchangeable, due to the MRF term of the prior for $\boldsymbol{C}$, and the discrete measure representation using $Q$ is no longer possible. 

\textbf{Prior coclustering properties under Pose}. A key distinction exists between common atom and shared atom models. As discussed in \cite{Dangelo2023}, the use of an infinite set of atoms with a common order across random measures together with SB priors induces complex dependencies among the observational weights in common atom models. This implies high prior coclustering probabilities, which might negatively impact the results a posteriori. This issue also arises in our setting: it is instructive to study how the prior coclustering probabilities between two entries $\prob{\theta_{i,j} = \theta_{i',j'}}$ changes under $\bm{\omega}_k \sim \Dirichlet_L(\boldsymbol{b}_0)$, i.e. under the \emph{shared atoms} model, instead of the \emph{common atoms} that a DP would induce. 
Proposition 1 of \cite{Lu2020} provides the predictive distribution of the BNP-MRF process for general Gibbs-type priors. Specialized to the DP and the case of two observations, we can derive the probability of ties between two groups, equal to $\prob{G_j=G_{j'}} = {e^{\beta}}/\left(\alpha+e^{\beta}\right)$. We can then prove the following proposition. The proof is deferred to Section~\ref{supp:proofs} of the Supplementary Material.
%\vspace{-1cm}
\clearpage
\begin{proposition}
\label{prop}
    The prior coclustering probability between two entries according to the shared atoms Pose model in~\eqref{eq:se_clusterassignment} with $\bm{\omega}_k \sim \Dirichlet_L(\boldsymbol{b}_0)$ for all $k$ and $\boldsymbol{\pi}\sim GEM(\alpha)$ is
    \begin{equation*}
        \prob{\theta_{i,j} = \theta_{i',j'}} 
    = \frac{1}{\alpha+e^{\beta}} \left[ e^{\beta}\left(\frac{1+b_0}{1+Lb_0}\right)^{1-\mathds{1}_{i=i'}} + \frac{\alpha}{L}\right].
    \end{equation*}
    In the same setting, if instead we assume $\boldsymbol{\omega}_k\sim GEM(\nu)$ for all $k$, we have
    \begin{align*}
    \prob{\theta_{i,j} = \theta_{i',j'}} 
    &= \frac{1}{\alpha+e^{\beta}} \left[ e^{\beta}\left(\frac{1}{1+\nu}\right)^{1-\mathds{1}_{i=i'}} + \frac{\alpha}{2\nu+1}\right].
    \end{align*}
\end{proposition}
A graphical depiction of these probabilities is shown in Supplementary Figures~\ref{fig:cocl1}–\ref{fig:cocl3}. Consistent with previous findings \cite{Dangelo2023}, the prior coclustering probability is higher under the common atoms model, highlighting its stronger a priori grouping and risk of spurious coclustering. Distinguishing between same-row ($i = i'$) and different-row ($i \neq i'$) entries is important, as when two groups (columns $j$ and $j'$) are clustered, their rows automatically share the same partition. The difference between models is most apparent for different rows, where clustering behavior depends on the observational weight ($b_0$ for shared atoms, $\nu$ for common atoms). For same-row pairs, coclustering under shared atoms is independent of $b_0$, but still depends on $\nu$ for common atoms. As the spatial regularization parameter $\beta$ increases, model differences diminish and clustering probabilities converge.

\subsection{The multiple-dataset case: Poseidon}
Given the definition of Pose for a single molecule dataset, we can simply extend the biclustering framework to multiple datasets.
Consider again the collection of $T$ datasets $\boldsymbol{\mathcal{Y}}=\{\boldsymbol{Y}^{(t)}\}_{t=1}^T$, each characterized by $N^{(t)}$ rows and $J$ columns. We now consider $T$ collections of row partitions $\boldsymbol{\mathcal{R}}^{(t)} = \{\boldsymbol{R}^{(t)}_k\}_{k=1}^K$ (one for each dataset $t=1,\ldots, T$), with a dataset-specific row partition $\boldsymbol{R}^{(t)}_k$ for each CC $k$. We also denote with $\boldsymbol{\theta}^{*(t)}$ the collection of cluster-specific parameters for each dataset $t = 1, \ldots,T$, and let $\boldsymbol{\theta}^{*} = (\boldsymbol{\theta}^{*(1)},\ldots, \boldsymbol{\theta}^{*(T)})$. Finally, we assume the existence of one partition of the columns $\boldsymbol{C}=\left(C_1,\ldots, C_J\right)$, shared by all the datasets, resulting in a common image segmentation.
Given these partitions, we assume the measurements are conditionally independent across datasets, setting $p(\boldsymbol{\mathcal{Y}}\mid \{\boldsymbol{\mathcal{R}}^{(t)}\}_{t=1}^{T}, \boldsymbol{C},\boldsymbol{\theta}^*) = \prod_{t=1}^T p(\boldsymbol{Y}^{(t)}\mid \boldsymbol{\mathcal{R}}^{(t)}, \boldsymbol{C},\boldsymbol{\theta}^{*(t)}).$ Then, we can express the model in the following form: %, useful for posterior inference:
\begin{equation}
\begin{gathered}
    \theta_{l}^{*(t)}  \simiid H^{(t)}, \quad t = 1, \ldots ,T \\
    \boldsymbol{C} \vert \boldsymbol{\pi},\beta \sim \textrm{BNP-MRF}(\boldsymbol{\pi},\beta),\quad\quad 
    R_{i,k}^{(t)} \vert \boldsymbol{\omega}^{(t)}_k \sim \textrm{Cat}_L(\boldsymbol{\omega}^{(t)}_k), \quad t = 1, \ldots ,T\\
    y_{i,j}^{(t)} \vert C_j, \{R^{(t)}_{i,k}\}_k, \{ \theta^{*(t)}_l \}_l  \simind f( y_{i,j}^{(t)} \vert \theta^{*(t)}_{R^{(t)}_{i,C_j}} ), \quad t = 1, \ldots ,T, %, \, i = 1,..,N^{(t)}, \, j=,..,J \\
\end{gathered}
\label{eq:poseidon}
\end{equation}
with $\boldsymbol{\pi} \sim GEM(\alpha),$ 
$\boldsymbol{\omega}^{(t)}_k \sim Dir_L(\boldsymbol{b}_0),\:\forall t,k,$ and each $H^{(t)}$ is a diffuse base measure. One can also specify a prior for the concentration parameter of the distributional DP, e.g., $\alpha \sim \mathrm{Gamma}(a_\alpha, b_\alpha)$, and a prior for the inverse temperature, e.g., $\beta\sim\mathcal{U}(0,B)$. %for the concentration parameter of the distributional DP. 
We call this model \emph{Pose for the Integration of Different Omics Nested data} (Poseidon).

Thanks to the hierarchical structure used in defining Poseidon, the model has the flexibility to model the molecule-specific datasets $\boldsymbol{Y}^{(t)}$, using dataset-specific parameters $\boldsymbol{\theta}^{*(t)}$; nevertheless, it allows for sharing information across these datasets to infer the shared partition of the columns, leveraging the signal provided the different molecules. In Section~\ref{sec:simulation}, we will explore these properties further with a simulation study. 
For our application, we will specialize model~\eqref{eq:poseidon} to fit a mixture of Gaussian distributions. As the first step, we need to transform the positive measurements of the original abundance matrix, here denoted with $\boldsymbol{Z}$, into real values as 
\begin{equation}
y_{i,j} = \Phi^{-1}\left(\frac{z_{i,j}-\min_{i,j}(z_{i,j})}{\max_{i,j}(z_{i,j})-\min_{i,j}(z_{i,j})}\right), \quad\text{for}\quad i=1,\ldots,N;\quad j=1,\ldots,J,
\label{eq:transform}
\end{equation}
where $\Phi^{-1}\left(\cdot\right)$ represents the inverse c.d.f. of a Normal distribution.
To complete our model specification, exploiting conjugacy, we finally specify Normal-Inverse Gamma priors for the base measures $H^{(t)}$, $t=1,\ldots,T$. Thus, for all $l$, we assume $\theta_{l}^{*(t)}=( \mu_{l}^{*(t)},\sigma_{l}^{2*(t)})$ where $\mu_l^{*(t)}\mid \sigma_{l}^{2*(t)} \sim \textrm{N}(m_0^{(t)}, \sigma_{l}^{2*(t)}/k^{(t)}_0)$ and $\sigma_{l}^{2*(t)}\sim \textrm{IG}(c^{(t)}_0,d^{(t)}_0)$. 
Recommended values for the hyperparameters are $(m^{(t)}_0, k^{(t)}_0, c^{(t)}_0, d^{(t)}_0) = \left(0, 0.1, 3, 2\right)$. We also set $\boldsymbol{b}_0 = 0.0001 \cdot \boldsymbol{1}_L$ to ensure sparse finite mixtures over the rows~\citep{MalsinerWalli2016}, $s_1 = s_2 = 1$, and $B=2$.

\section{Posterior inference via variational Bayes}\label{sec:VB}

We perform posterior inference using a tailored variational Bayes (VB) algorithm to grant an efficient estimation of Poseidon on our high-dimensional MALDI-MSI data. In short, let $\boldsymbol{\Theta}=\left(\{\boldsymbol{\mathcal{R}}^{(t)}\}_{t=1}^{T},\boldsymbol{C},\boldsymbol{\theta},\boldsymbol{\pi},\boldsymbol{\omega},\alpha,\beta\right)$ represent the set of all the parameters in our model. 
After a family of approximating densities $\mathcal{Q}$ is specified over $\boldsymbol{\Theta}$, we seek to find the member $q^*(\boldsymbol{\Theta})$ of $\mathcal{Q}$ that minimizes the Kullback-Leibler divergence ($\mathcal{D}_{KL}$) with true posterior distribution. Note that minimizing this divergence is equivalent to maximizing the Evidence Lower BOund (ELBO), defined as $ELBO(q) = \mathbb{E}\left[ \log p\left(\boldsymbol{\Theta},\boldsymbol{Y}\right)\right]-\mathbb{E}\left[\log q\left(\boldsymbol{\Theta}\right)\right]$. The variational family $\mathcal{Q}$ is typically characterized by variational parameters $\boldsymbol{\lambda}$. The optimization problem is thus reduced to finding the optimal parameters $\boldsymbol{\lambda}^{*} = \arg\min_{\boldsymbol{\lambda}} \operatorname{\mathcal{D}_{KL}}\left( q_{\boldsymbol{\lambda}}(\boldsymbol{\Theta}) \parallel p(\boldsymbol{\Theta} \mid \mathbf{Y}) \right).$ We opt for a mean-field variational approximation, which chooses $\mathcal{Q}$ assuming the parameters are mutually independent. To find $\boldsymbol{\lambda}^*$, we resort to a Coordinate Ascent Variational Inference (CAVI) procedure, detailed in Algorithm~\ref{algo::caviposeidon}. The single-dataset case (Pose) is easily recovered when $T=1$.
Almost all the CAVI steps can be performed in closed form, as the corresponding full conditionals are known distributions. The only steps that require approximations are the updates for the parameters of the distributions of $v_k$'s and the inverse temperature $\beta$. For the former case, we mimic \cite{Lu2020} and assume $\beta=0$. In the latter case, we proceed as delineated in Section~\ref{supp:estimating_beta} of the Supplementary Material.

To fit the model, we initialize Algorithm~\ref{algo::caviposeidon} using different starting points, stopping the computations when the relative increment in the ELBO is below a certain threshold, say, for example, 0.001\%. Once all the runs are performed, we retain the estimate that obtained the highest value of the ELBO. The details for the computation of the ELBO components are in Section~\ref{supp:elboposeid} of the Supplementary Material.

\begin{algorithm}[H]
\SetAlgoLined
{\footnotesize %for submission, it was \small
\hspace*{0em}\textbf{Input:} $r \gets 0$. Randomly initialize $\boldsymbol{\lambda}^{(0)}$. Define the error bound $\epsilon$ and randomly set $\Delta>\epsilon$. 

\While{$\Delta(r-1,r) > \varepsilon$}{
    Let $\boldsymbol{\lambda} = \boldsymbol{\lambda}^{(r)};$  Set $r = r+1$ ; \\
    Update the variational parameters according to the following CAVI steps:
    \begin{enumerate}
    \item For $j=1,\dots,J$, $q^\star(C_{j})$ is a $K$-dimensional multinomial, with $q^\star(C_{j}=k)=\rho_{j,k}$ for $k=1,\dots,K$, and
    $$
       \log\rho_{j,k} = 
            g(\bar{a}_k,\bar{b}_k) + \sum_{r=1}^{k-1}g(\bar{b}_r,\bar{a}_r)
       + \sum_{l=1}^L \sum_{t=1}^{T}\sum_{i=1}^{N} \xi^{(t)}_{i,k,l}\ell^{(t)}_{i,j,l}+\beta \sum_{q\in \mathcal{N}_j}\rho_{q,k},
    $$
     where 
    $g(x,y) = \psi(x) - \psi(x+y)$, with $\psi$ being the digamma function, and $$\ell^{(t)}_{i,j,l} = -\frac12 \left[ \log(d^{(t)}_l) - \psi(c^{(t)}_l) +\frac{1}{k^{(t)}_l}+\frac{c^{(t)}_l}{d^{(t)}_l} (y^{(t)}_{i,j}-m^{(t)}_l)^2\right].$$
    \item  For $k=1,\dots,K$, $t=1,\ldots,T$ and $i=1,\dots,N$, $q^\star(R_{i,k})$ is a $L$-dimensional multinomial,\\with $q^\star(r^{(t)}_{i,k}=l)=\xi^{(t)}_{i,j,l}$ for $l=1,\dots,L$, and
    \begin{align*}
       \log\xi^{(t)}_{i,k,l} = \sum_{j=1}^J \rho_{j,k} \ell^{(t)}_{i,j,l}+  h_{l}(\boldsymbol{p}^{(t)}_k),
    \end{align*}
    with $M_k=\sum_{j=1}^J \rho_{j,k}$, $\ell^{(t)}_{i,j,l}$ is defined as in Step (1), and $h_{l}(\boldsymbol{p}^{(t)}_k) = \psi(p^{(t)}_{l,k})-\psi(\sum_{l=1}^L p^{(t)}_{l,k})$.
    \item  For $k=1,\dots,K$ and $t=1,\ldots,T$, {$q^\star(\boldsymbol{\omega}^{(t)}_k)$} is $\Dirichlet_L(\boldsymbol{p}^{(t)}_k)$ with
    $ p^{(t)}_{l,k} = b^{(t)}_{l,k} + \sum_{i=1}^{N} \xi^{(t)}_{i,k,l}.$
    \item  %q^\star(\bpi)$ is $\Dirichlet_K(\tilde{\boldsymbol{p}})$ with $\tilde{p}_k = a_k + \sum_{j=1}^J \rho_{j,k}$ for $k=1,\dots,K$.\textcolor{red}{update this - manca un pezzo in Lu, e anche da noi - JUNE: mi chiedo, se $\beta$ e' parametro fisso ora, c'e ancora termine mancante o posso ignorarlo?}\\
    For $k=1,\dots,T$, $q^\star(v_{k})$ is approximated as in~\cite{Lu2020} with a $\Beta(\bar{a}_k,\bar{b}_k)$ distribution with $$
        \bar{a}_{k} = 1 + \sum_{j=1}^J\rho_{j,k}, \quad \bar{b}_{k} = s_1/s_2 + \sum_{j=1}^J\sum_{q=k+1}^{K-1}\rho_{j,q}.$$
 
    \item   For $l=1,\dots,L$ and $t=1,\ldots,T$, {$q^\star(\theta^{*(t)}_l)$} is a $NIG(m^{(t)}_l,k^{(t)}_l,c^{(t)}_l,d^{(t)}_l)$ distribution with parameters\\$k^{(t)}_l = n_l^{(t)} + \kappa_0$, $c^{(t)}_l = \tau_0 +n_l^{(t)}/2,$
    \begin{equation*}
        m^{(t)}_l = \frac{\kappa_0\:m_0+n_l^{(t)}\bar{y}^{(t)}_l}{n_l^{(t)}+\kappa_0}, \quad
        d^{(t)}_l = \gamma_0 +\frac{n_l^{(t)} \kappa_0}{2 k^{(t)}_l}\left(\bar{y}^{(t)}_l-m_0\right)^2 + \frac12 \sum_{j=1}^J\sum_{i=1}^{N^{(t)}} Q^{(t)}_{i,j,l}\left(y^{(t)}_{i,j}-\bar{y}^{(t)}_l\right)^2,
    \end{equation*}
    where $Q^{(t)}_{i,j,l} = \sum_{k=1}^{K} \xi_{i,k,l}^{(t)}\rho_{j,k}$,  $n_l^{(t)} = \sum_{i=1}^{N^{(t)}}\sum_{j=1}^J Q^{(t)}_{i,j,l}$, and $\bar{y}^{(t)}_l = 
( \sum_{i=1}^{N^{(t)}}\sum_{j=1}^J Q^{(t)}_{i,j,l}y_{i,j}) / n_l^{(t)}$.

  \item  $q^\star(\alpha)$ is a $\mathrm{Gamma}(s_1,s_2)$ distribution with parameters
    $$s_1 = a_\alpha + K-1, \quad s_2 = b_\alpha - \sum_{k=1}^{K-1} g(\bar{b}_{k},\bar{a}_{k}).$$

    \item Find an updated value for the expected value of the inverse temperature $\beta$ following the approximating procedure reported in Section~\ref{supp:estimating_beta} of the Supplementary Material.

\end{enumerate}
    
    Compute $\Delta(r-1,r) = \left[\mathrm{ELBO}(\boldsymbol{\lambda}^{(r)})- \mathrm{ELBO}(\boldsymbol{\lambda}^{(r-1)})\right]/\mathrm{ELBO}(\boldsymbol{\lambda}^{(r-1)})$.\\ 
 }
 \Return{$\boldsymbol{\lambda}$}
 }
 \caption{CAVI updates for the POSEIDON model}
\label{algo::caviposeidon}
\end{algorithm}

Similarly to the case of the nested model presented in \cite{Dangelo2023}, resorting to a variational Bayes approach not only provides estimates in a timely and efficient manner but also addresses the problem of label switching and, more importantly, offers a direct method for estimating the partition of rows and columns. Indeed, once the algorithm has converged, we are left with two sets of posterior variational clustering probabilities that we denote with $\hat{\rho}_{j,k}=q^*(C_j=k)$ and $\hat{\xi}_{i,l,k}=q^*(R_{i,k}=l)$. The former indicates how likely the $j$-th column is to be assigned to the $k$-th cluster. Similarly, the latter represents the probability that the $i$-th row belongs to the $l$-th cluster within the $k$-th CC.
Thus, we estimate the pixel segmentation as
%\begin{equation}
$\hat{C}_j=\arg\max_{k=1,\ldots,K}\hat{\rho}_{j,k}\:\: \forall j.$
%\end{equation}
Analogously, the RC solutions nested in the $k$-th column cluster are given by
%\begin{equation}
$\hat{R}_{i,k}=\arg\max_{l=1,\ldots,L}\hat{\xi}_{i,k,l}\:\: \forall i,k.$
%\end{equation}
%
It is important to note that, with this procedure, we obtain $K$ partitions of the rows, one for each \emph{potential} CC. However, only the ones corresponding to non-empty CCs are interesting.

\section{Simulations} \label{sec:simulation}
\subsection{Biclustering recovery}
\label{sec:simulation1}
One of the key features that distinguishes Pose from most biclustering methods is its ability to recover different row-clustering structures within different CCs.
In contrast, most block-clustering approaches \citep[see, e.g.,][]{tan2014sparse, govaert2013co} assume a single row-clustering structure shared across all CCs.
This flexibility is particularly important for MALDI-MSI data, where we expect peaks to occur at different \textit{m/z} (rows) in different CCs, making traditional block-clustering methods poorly suited to the task.

To assess Pose’s performance in recovering biclustering structure, we use synthetic datasets designed to mimic the characteristics of MALDI data.
In these datasets, certain \textit{m/z} values appear as peaks with high activation levels (mean = 1.5), their immediate neighbors show medium activation (mean = 0.75), and all other rows are treated as noise (mean = 0).
Peak locations vary across CCs, and we simulate a moderately high noise level (sd = 1).
We generate 30 synthetic datasets sharing the same column partition.
Further details on the data generation process are provided in Section~\ref{supp_sec_biclrecovery} of the Supplementary Material.

\begin{figure}[h!]
	\centering
	\includegraphics[width=\linewidth]{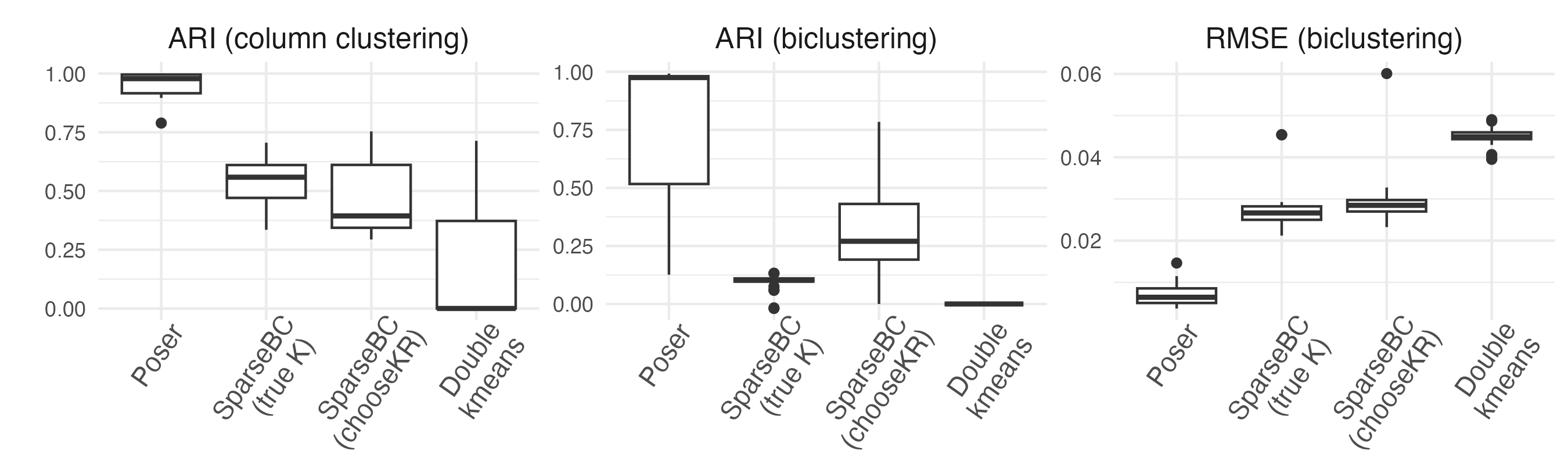}
	\caption{Recovery of column partition (left panel) and biclustering structure (central and right panel), measured by ARI (left and center) and RMSE (right) in synthetic data.}
	\label{fig:simB_results}
\end{figure}

We compare the biclustering performance of:
(a) Pose (fixing at $\beta = 1$),
(b) \texttt{sparseBC} \citep{tan2014sparse} with the true number of CCs $K$ (\texttt{true K}),
(c) \texttt{sparseBC} with $K$ selected via its internal procedure (\texttt{chooseKR}), and
(d) \texttt{Double k-means}, which first applies k-means to columns and then, within each CC, applies k-means to the corresponding rows.
Figure~\ref{fig:simB_results} shows recovery of the true CC and biclustering structure, measured by the Adjusted Rand Index (ARI), as well as the root mean squared error (RMSE) in recovering the biclustering matrix.
Overall, Pose delivers the best performance, with median ARI values close to 1 and the lowest RMSE.
\texttt{sparseBC} performs substantially worse, with a median ARI of around 0.5 for CC recovery and larger RMSE values, reflecting the limitations of its block-clustering assumption.
Surprisingly, \texttt{Double k-means}, despite its greater flexibility compared to \texttt{sparseBC}, yields the poorest results.

% Note: the simulations are not run with beta = 0, but beta = 1.

\subsection{Informative datasets guide the shared clustering estimation}
\label{sec:simulation2}

A desired feature of Poseidon is that the shared column partition should be mostly driven by datasets containing informative observations rather than being negatively impacted by those potentially dominated by noise. To investigate this property, we compare the performance of Pose and Poseidon models without spatial interaction. 

We consider five datasets, denoted ${D}^{(r)}$, each with $J=20$ columns and $N^{(r)}=50$ rows, for $r = 0, \dots, 4$, assuming a common biclustering structure. Two CCs are defined. In the first one, the top half of the rows have entries sampled from a $N(\mu^{(r)}, 1)$ distribution, with $\mu^{(0)}=4$ and $\mu^{(r)}=1/r$ for $r = 1, \dots, 4$, while the remaining rows follow a $N(0, 1)$ distribution. The second CC contains a unique RC, with all entries sampled from $N(0, 1)$  distribution. The magnitude of $\mu^{(r)}$ controls the signal-to-noise ratio: as $r$ grows, the two RC distributions increasingly overlap and noise dominates, complicating the recovery of the true partitions.
More details about the data-generating process and hyperparameter specifications are reported in Section~\ref{supp:sec_simu_D0D4} of the Supplementary Material.

The results are summarized in the top two rows of Table~\ref{tab:DCARI}, where we report the averages (and std. dev.) over 100 replicas of the ARIs computed between the ground truth and the estimated column and row partitions, respectively. While both partitions are largely recovered for ${D}^{(0)}$ and ${D}^{(1)}$, performance deteriorates sharply for the remaining datasets as the signal-to-noise ratio decreases. We then estimate a multi-dataset model with $T = 2$, considering four ``collections” of datasets formed by combining the reference dataset with each noisy dataset: $\{D^{(0)}, D^{(r)}\}$ for $r = 1, \ldots, 4$. The results, summarized in the central rows of Table~\ref{tab:DCARI}, report the ARI indexes for the CCs and the two distinct RCs estimated on the reference and noisy datasets, respectively.
First, we note that including $D^{(0)}$ drives the shared CC solution, effectively mitigating the noise in the other datasets. Second, the estimation of the RCs for the noisy datasets shows significant improvement compared to the results from single-dataset models without affecting the results on the reference dataset. Thus, the multi-dataset approach has an advantage in recovering the underlying structure. 

The observed improvement raises an important question. Since we are working with multiple datasets sharing the same number of columns, one may ask whether similar results can be obtained with a simpler strategy: applying the single-dataset Pose model to tables constructed by row-stacking the reference dataset with one of the noisy datasets. We denote these datasets as $\left[D^{(0)}, D^{(r)}\right]$, for $r = 1, \ldots, 4$. 
The clustering performances are summarized in the last two rows of Table~\ref{tab:DCARI}. We observe that the CC solutions exhibit robustness, albeit with a slight underperformance compared to those obtained with Poseidon. In contrast, contamination affects the joint RC configurations. As expected, RC estimation improves compared to the one on the individual noisy dataset. Still, it comes at the cost of compromising the recovery of the true partition in the reference dataset, leading to the creation of additional RCs, as detailed in Supplementary Table~\ref{supp:tab:NCLUST_RC}.
\begin{table}[h]
    \centering
    \begin{tabular}{lcccccccccc}
    \toprule
     Pose & \multicolumn{2}{c}{${D}^{(1)}$} & \multicolumn{2}{c}{${D}^{(2)}$} & \multicolumn{2}{c}{${D}^{(3)}$} & \multicolumn{2}{c}{${D}^{(4)}$}\\
    \cmidrule(lr){1-1} \cmidrule(lr){2-3} \cmidrule(lr){4-5} \cmidrule(lr){6-7} \cmidrule(lr){8-9} 
    CC &  0.948 & (0.101) & 0.437 & (0.447) & 0.008 & (0.080) & 0.000 & (0.000)\\
    RC &  0.720 & (0.210) & 0.313 & (0.337) & 0.006 & (0.058) & -0.002 & (0.010)\\
    \midrule
      Poseidon & \multicolumn{2}{c}{$\{D^{(0)},D^{(1)}\}$} & \multicolumn{2}{c}{$\{D^{(0)},D^{(2)}\}$} & \multicolumn{2}{c}{$\{D^{(0)},D^{(3)}\}$} & \multicolumn{2}{c}{$\{D^{(0)},D^{(4)}\}$}\\
    \cmidrule(lr){1-1} \cmidrule(lr){2-3} \cmidrule(lr){4-5} \cmidrule(lr){6-7} \cmidrule(lr){8-9} 
   CC &  0.991 & (0.045) & 0.996 & (0.028) & 0.990 & (0.035) & 0.993 & (0.032)\\
    RC Ref. & 0.970 & (0.115) & 0.966 & (0.123) & 0.977 & (0.095) & 0.967 & (0.119)\\
    RC Noisy & 0.802 & (0.176) & 0.673 & (0.147) & 0.360 & (0.190) & 0.136 & (0.130)\\
        \midrule
      Stacked & \multicolumn{2}{c}{$[D^{(0)},D^{(1)}]$} & \multicolumn{2}{c}{$[D^{(0)},D^{(2)}]$} & \multicolumn{2}{c}{$[D^{(0)},D^{(3)}]$} & \multicolumn{2}{c}{$[D^{(0)},D^{(4)}]$}\\
    \cmidrule(lr){1-1} \cmidrule(lr){2-3} \cmidrule(lr){4-5} \cmidrule(lr){6-7} \cmidrule(lr){8-9} 
    CC & 0.981 & (0.057) & 0.981 & (0.056) & 0.978 & (0.060) & 0.976 & (0.066)\\
    RC Aggr. & 0.767 & (0.172) & 0.681 & (0.164) & 0.534 & (0.166) & 0.434 & (0.107)\\
    RC Ref.  & 0.875 & (0.191) & 0.839 & (0.196) & 0.803 & (0.204) & 0.824 & (0.219)\\
    RC Noisy & 0.771 & (0.174) & 0.620 & (0.163) & 0.339 & (0.2) & 0.094 & (0.134)\\
    \bottomrule
    \end{tabular}
    \vspace{1cm}
    \caption{Average and standard deviation of the adjusted Rand indexes obtained comparing the ground truth and the estimated column and row partitions on the noisy datasets using Pose, Poseidon, and stacking the reference and noisy datasets (in all models, $\beta=0$). For comparison, the indices for the RC and CC solutions estimated on the reference dataset $D^{(0)}$ using Pose are $0.992\:\:(0.037)$ and $0.914\:\:(0.181)$.}
    \label{tab:DCARI}
\end{table}

\section{ccRCC image segmentation and molecular signal extraction}

\label{sec:application}

We fit Poseidon to the ccRCC dataset described in Section~\ref{subsec:data} after applying the transformation in~\eqref{eq:transform}. To estimate the model, we set the upper bound for the number of CCs and RCs to $K=30$ and $L=40$, respectively. We run the algorithm using 200 different starting points on a Linux machine with 32 GB of memory. The average running time is 44.30 seconds, with a standard deviation of 13.22 seconds. 

\begin{figure}[t!]
    \centering
    \includegraphics[width=\linewidth]{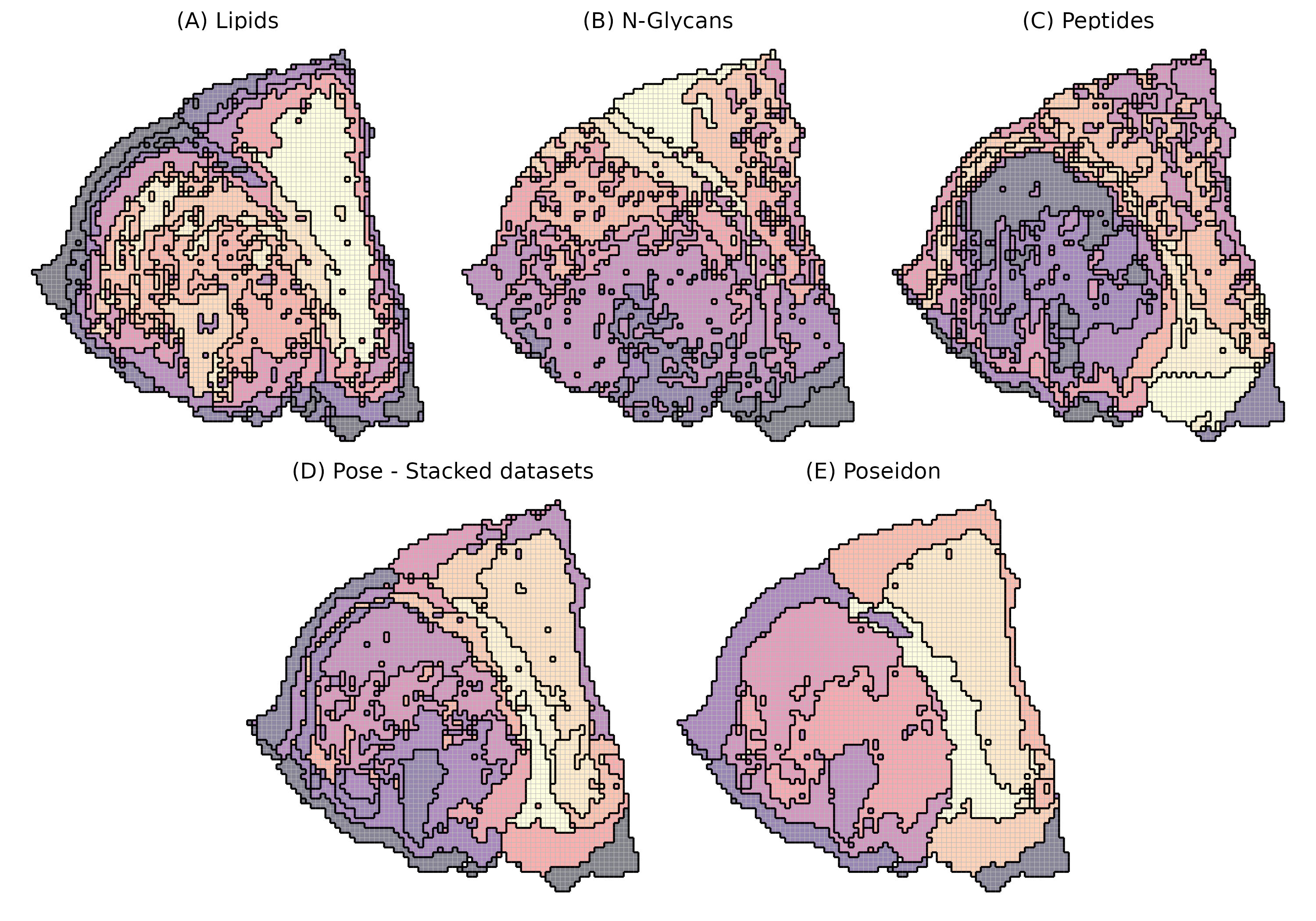}
    \caption{Tissue segmentation obtained with fitting single-dataset Pose models to the separate molecules (top panels, A--C) and the stacked datasets (bottom left panel, D). These results are compared with Poseidon's multi-dataset approach (bottom right panel, E).}
    \label{fig:fig4_res_clustering}
\end{figure}

\textbf{Tissue segmentation}. To evaluate the advantages of Poseidon over single-dataset approaches, we compare its performance against separate Pose models fitted to four distinct datasets: (1) lipids, (2) N-glycans, (3) peptides, and (4) a stacked dataset combining all three molecular classes. The resulting spatial partitions are reported in Figure~\ref{fig:fig4_res_clustering}.
Segmentation at the single-molecular level offers distinct insights into the biological architecture of histological tissue. N-glycans and peptides are closely linked, as N-glycans typically attach to specific peptide sequences, functioning as ``labels" that direct proteins to their appropriate cellular compartments. This post-translational modification ensures the proper localization and functionality of proteins. Given this biological relationship, one would expect similar spatial partitioning patterns for these two molecular layers. Indeed, in the healthy region of the tissue, the clustering of N-glycans and peptides shows a high degree of concordance. Conversely, within the tumor nodule, their clustering patterns diverge, likely reflecting disrupted glycosylation processes in cancer \citep{wallace2024n}.
In contrast to these two modalities, lipid-based segmentation reveals distinct clusters within the tumor nodule, highlighting regions infiltrated by inflammatory cells and areas with hemorrhagic features (see also Figure~\ref{suppfig:S1_ccrcc_legend}). Additionally, the lipid profile in the healthy region displays concentric stratification, possibly indicating subtle variations associated with early inflammatory processes. These observations are further reinforced by the segmentation obtained using a single-dataset Pose model applied to the stacked datasets, emphasizing the importance of a principled multiomics integration strategy.
In fact, when integrating all three molecular layers, Poseidon yields a more coherent and biologically meaningful tissue segmentation. By leveraging complementary information from different analytes, the model reduces the influence of noise and dataset-specific artifacts that can drive spurious clusters in single-data analyses. This integration yields a unified representation of the tissue's molecular landscape, capturing shared biological patterns across the various omics layers. Importantly, Poseidon reveals distinct multiomic alterations within both the tumor microenvironment and adjacent histologically normal tissue regions that remain undetected by standard histopathological examination (see Figure~\ref{fig:fig2_data}, panel A).

\begin{figure}[ht]
    \centering
    \includegraphics[width=\linewidth]{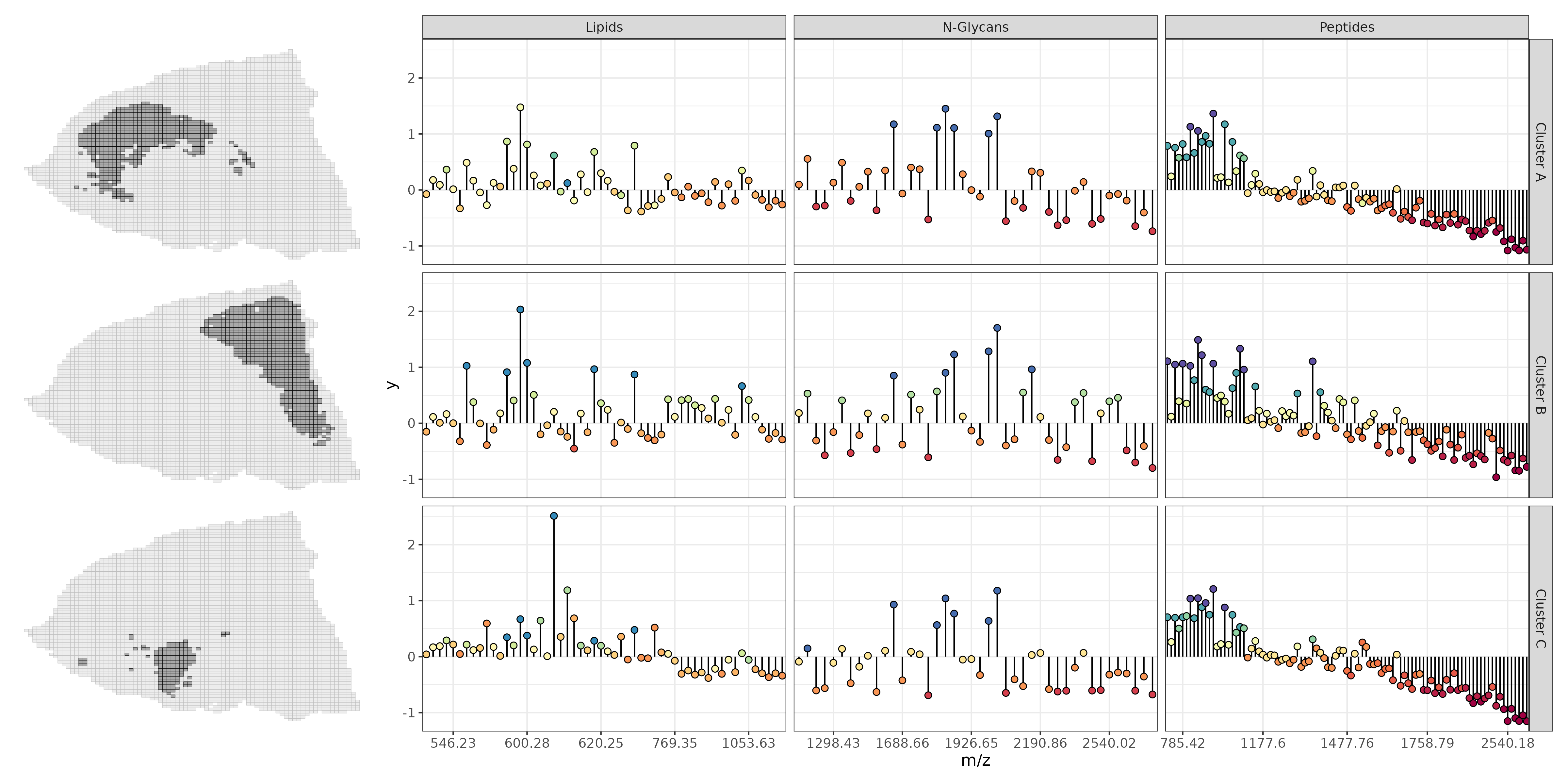}
    \caption{Spatial and spectral characterization of three representative regions of interest identified within the tissue. The left panels show the spatial distribution of each CC, highlighting the localization patterns across the sample. Right panels display the corresponding posterior averages of the analytes' rescaled abundances in each CC, segmented by molecular class: lipids, N-glycans, and peptides. Therefore, each spectrum illustrates the relative intensity of key \textit{m/z} features contributing to the cluster-specific molecular profile, and it is colored according to the RC assignments.}
    \label{fig:fig5_three_clusters}
\end{figure}

\textbf{Results on \textit{m/z} in specific CCs}. To highlight the type of inference enabled by the Poseidon formulation, we focus on the results obtained for three CCs associated with biologically meaningful tissue regions. These CCs are reported in the left panels of Figure~\ref{fig:fig5_three_clusters}. Two correspond to distinct areas within the tumor nodule, depicted in the top (Cluster A) and bottom (Cluster C) panels. 
More specifically, the region captured in Cluster A was annotated by the pathologist as a mixed-grade tumor, representing an intermediate stage between \emph {grade 2} and \emph{grade 3}. The region in Cluster C, instead, contains hemorrhagic tissue, where tumor cells were classified as \emph{grade 2}. Finally, the central panel (Cluster B) shows a healthy region of the kidney cortex. Distinguishing between the two tumor areas is challenging, even for experienced pathologists, due to their subtle morphological differences. Remarkably, Poseidon's multiomics integration enabled a spatially resolved characterization of molecular behavior across these regions, highlighting differences that are difficult to detect visually. More importantly, we can analyze the behavior of the \textit{m/z} values across the three highlighted clusters. The right panels of Figure~\ref{fig:fig5_three_clusters} display the posterior mean of each row assigned to the CCs. The dots, colored according to their \textit{m/z} partition, reveal how specific molecular signals contribute to each molecular class. By jointly considering these spatial and molecular patterns, we can identify the most influential signals that characterize each region.

The N-glycans display overall similarities across the tumor regions (Clusters A and C), consistent with the well-documented glycan alterations associated with cancer, as opposed to the distinct profile observed in the benign area (Cluster B). From a lipidomic perspective, we observe marked differences in the higher \textit{m/z} range, likely reflecting the divergent metabolic states between healthy and tumor tissues. Notably, certain small lipids within the 606–619 \textit{m/z} range show elevated mean abundance in Cluster C. This may be attributed to the presence of blood-derived lipids rather than tissue-specific ones, thus highlighting the hemorrhagic regions within the tumor nodule.
In contrast, peptides tend to exhibit higher abundance in the healthy region. Interestingly, several peptides also show increased levels in the tumor areas. Among these are peptides corresponding to nuclear proteins, such as \textit{m/z} 944.53 (\textit{histone H2A}) and \textit{m/z} 1116.56 (putatively identified as \textit{Heterochromatin Protein 1-binding protein 3}), which are recognized tumor markers in the MSI literature \citep{denti2024spatially}. These proteins are typically overexpressed in rapidly proliferating cells, such as those found in cancerous tumors. While the proteomic landscape in cancer has been extensively studied, the functional roles of N-glycans and lipids remain less explored. This is especially true when considering their spatial distribution across distinct tissue types \citep{vasseur2022lipids,wallace2024n}. Our findings offer a foundation for future investigations into the spatially resolved roles of these molecular layers in disease processes. Supplementary Section~\ref{supp:other_mol} provides additional examples of the spatial localization patterns for each molecular level.

\section{Discussion}
\label{sec:conclusions}
We introduced Poseidon, a BNP method for the joint analysis of multi-molecule MSI data, designed to effectively integrate the heterogeneous information provided by different spatially resolved multiomics experiments. By combining a Potts prior with a separate exchangeable clustering framework, we could simultaneously segment a tissue and cluster molecular features across omics classes while preserving spatial dependencies. This principled integration improved over both single-dataset and naively combined approaches, as demonstrated in our analysis of lipids, N-glycans, peptides, and stacked datasets. The improvements observed -- both in terms of biologically meaningful tissue segmentation and the discovery of plausible molecular biomarkers -- underscore the importance of integrated models for spatially structured molecular data.

This work opens up several promising research directions. First, one can consider different likelihood specifications, allowing for a more flexible modeling of the data without resorting to pre-processing transformations.
Second, alternative spatially informed priors could be explored as an alternative to the hidden MRF, which is known to pose computational challenges. For example, thresholded Gaussian processes or spatially dependent SB priors~\citep[e.g.,][]{linderman2015multinomial} may offer more flexible and scalable alternatives. Finally, incorporating stochastic variational inference could enable scalability to larger datasets, facilitating information sharing across multiple tissues from the same patient collected over time, or across tissues from different patients with the same disease. This would potentially allow for integrated, one-step differential analyses across space, time, and individuals.

%% file: 02_SUPP_skeleton.tex
\renewcommand{\thesection}{S.\arabic{section}}
\renewcommand{\theequation}{S.\arabic{equation}}
\setcounter{figure}{0}
\setcounter{table}{0}
\setcounter{equation}{0}
\setcounter{section}{0}
\renewcommand{\figurename}{Fig.}
\renewcommand{\thefigure}{S.\arabic{figure}}
\renewcommand{\tablename}{Tab.}
\renewcommand{\thetable}{S.\arabic{table}}

\section{Additional details about the MALDI-MSI data used in the analysis}
\label{supp:sec_rawdata}
\begin{figure}[h!]
    \centering
    \includegraphics[width=1\linewidth]{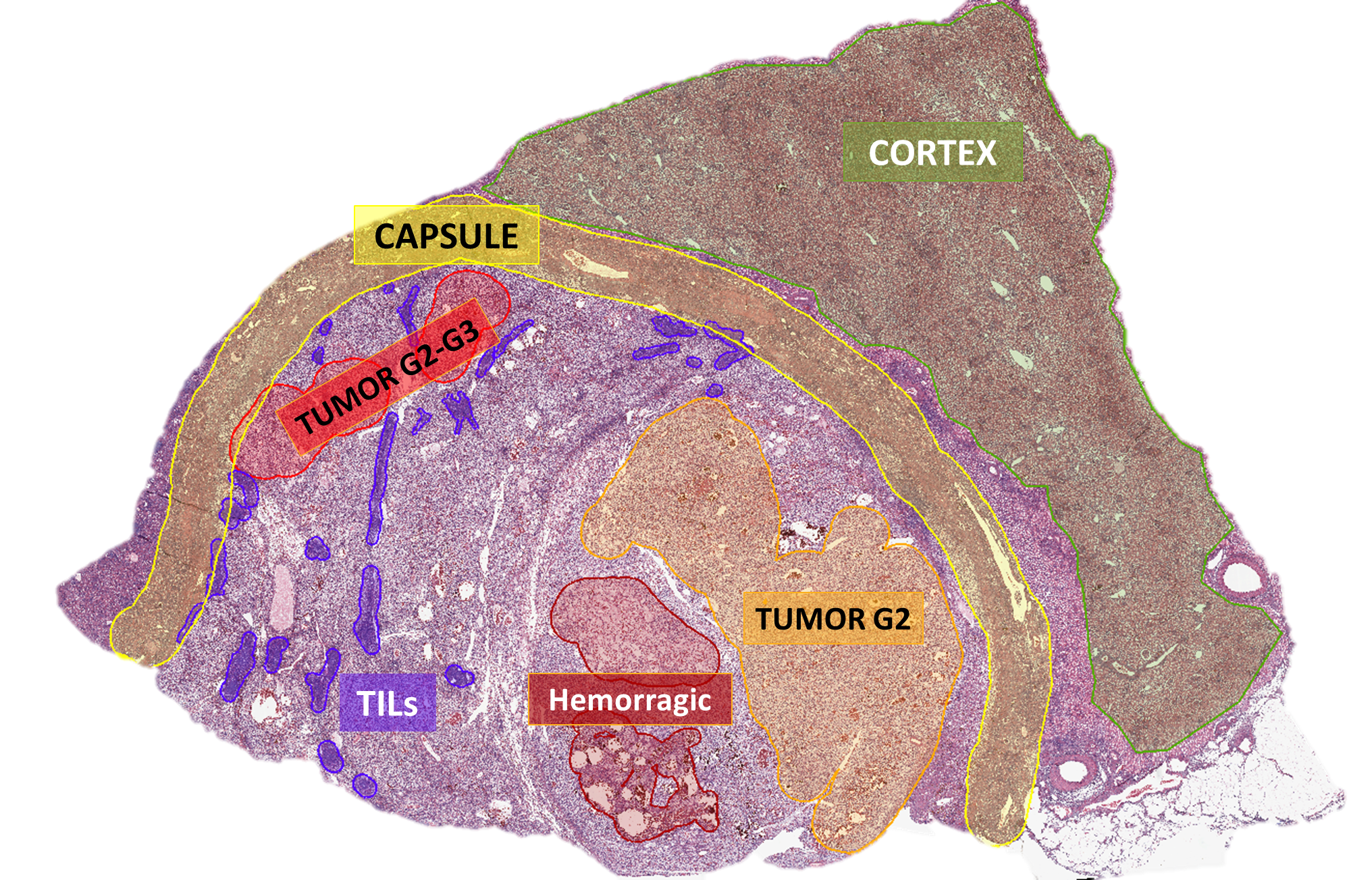}
    \caption{Annotated representation of the H\&E-stained ccRCC tissue used for the analysis, with regions of interest defined by an expert pathologist: green indicates healthy cortex; yellow, the capsule surrounding the tumor nodule; red, a tumor region with mixed grades (\textit{grade 2} and \textit{grade 3}); orange, a \textit{grade 2} tumor region; brown, a hemorrhagic area; and purple, tumor-infiltrating lymphocytes (TILs).}
    \label{suppfig:S1_ccrcc_legend}
\end{figure}
\noindent
Figure~\ref{suppfig:S1_ccrcc_legend} shows an annotated view of the H\&E-stained clear cell Renal Cell Carcinoma (ccRCC) tissue we analyzed. This image is a high-resolution version of the one reported in panel (A) of Figure~\ref{fig:fig2_data} of the main paper. In this detailed representation, regions of interest identified by an expert pathologist are color-coded as follows: green for healthy cortex, yellow for the capsule surrounding the tumor nodule, red for a tumor region with mixed grades (\textit{grade 2} and \textit{grade 3}), orange for a \textit{grade 2} tumor region, brown for a hemorrhagic area, and purple for tumor-infiltrating lymphocytes (TILs). From this tissue, raw data were extracted and prepared for the statistical analysis using a common statistical pipeline~\citep{Capitoli2024} as MALDI-MSI datasets require several pre-processing steps to reduce physiological, instrumental, and analytical variability. 
\clearpage
The pre-processing steps can be summarized as follows:

\begin{enumerate}
\item \textbf{Baseline subtraction:} Estimating and removing electrical noise and chemical impurities from the spectra.
\item \textbf{Smoothing:} Reducing interfering peaks from sources unrelated to the patient sample.
\item \textbf{Normalization:} Bringing all spectra to a common intensity range.
\item \textbf{Alignment:} Correcting slight differences in \emph{m/z} values to ensure the same signal is identified across spectra.
\item \textbf{Peak detection:} Retaining only signals with absolute intensity above the signal-to-noise ratio.
\end{enumerate}

\clearpage
\section{Additional details about the methodology and the algorithm}
\subsection{Proofs of Proposition~\ref{prop}}
\label{supp:proofs}
We aim to derive the expressions for the observational coclustering probability between two data points under our nested prior, i.e., $\prob{\theta_{i,j} = \theta_{i',j'}}$. Recall that $\bm{\omega}_k \sim \Dirichlet_L(b_0,\dots,b_0)$ and $\boldsymbol{\pi}\sim \textrm{BNP-MRF}(\boldsymbol{\pi};\beta)$ with $\boldsymbol{\pi}\sim \textrm{GEM}(\alpha)$.
First, let $\varrho_{j,j'}=\prob{G_j=G_{j'}}$. Then, we have
\begin{align*}
    \prob{\theta_{i,j} = \theta_{i',j'}} 
    &= \E{\prob{\theta_{i,j} = \theta_{i',j'}\mid G_j,G_{j'}}} \\
    &= \mathbb{E}\left[\prob{\theta_{i,j} = \theta_{i',j'}\mid G_j=G_{j'}} \varrho_{j,j'}\right] + \mathbb{E}\left[ \prob{\theta_{i,j} = \theta_{i',j'}\mid G_j\neq G_{j'}} (1-\varrho_{j,j'}) \right]
\end{align*}

In case of two samples, we can obtain $\varrho_{j,j'}=\prob{G_j=G_{j'}} =e^\beta/(\alpha+e^\beta)$ from the predictive distribution presented in~\citet{Lu2020}.
The second term in the summation is identical to the partially exchangeable case, see~\cite{Dangelo2023,Denti2021}. However, under the separate exchangeability, if column $j$ and $j'$ are clustered together, then the rows must share the same partition. Therefore,

\begin{align*}
    \prob{\theta_{i,j} = \theta_{i',j'}\mid G_j=G_{j'}}&=
    \mathds{1}_{\{i=i'\}}\prob{\theta_{i,j} = \theta_{i,j'}\mid G_j=G_{j'}}+
    \mathds{1}_{\{i\neq i'\}}\prob{\theta_{i,j} = \theta_{i',j'}\mid G_j=G_{j'}}\\&=
     \mathds{1}_{\{i=i'\}}\cdot 1+
    \mathds{1}_{\{i\neq i'\}}\sum_{l=1}^L\prob{\theta_{i,j} = \theta^*_l, \:\theta_{i',j'}=\theta^*_l \mid G_j=G_{j'}}\\&=
          \mathds{1}_{\{i=i'\}} + \mathds{1}_{\{i\neq i'\}}\sum_{l=1}^L \omega_{l,j}^2.
\end{align*}

Therefore,

\begin{align*}    
\prob{\theta_{i,j} = \theta_{i',j'}} 
    &= \frac{e^\beta}{e^\beta+\alpha}\, \left(\mathds{1}_{\{i=i'\}} + \mathds{1}_{\{i\neq i'\}}\E{\sum_{l=1}^L \omega_{l,j}^2} \right)+ \frac{\alpha}{e^\beta+\alpha} \E{\sum_{l=1}^L \omega_{l,j} \,\omega_{l,j'}}\\
    & \overset{(\star)}{=} \frac{e^\beta}{(e^\beta+\alpha)} \left[ \left(\frac{1+b_0}{1+Lb_0}\right)^{1-\mathds{1}_{\{i=i'\}}} + \frac{\alpha}{L}\right],
\end{align*}
where in step $(\star)$ we used
\begin{equation*}
    \E{\sum_{l=1}^L \omega_{l,j}^2} = \sum_{l=1}^L \E{ \omega_{l,j}^2} = \sum_{l=1}^L \frac{(1+b)}{L(1+Lb)} = \frac{1+b_0}{1+Lb_0},
\end{equation*}
and
\begin{equation*}
    \E{\sum_{l=1}^L \omega_{l,j} \, \omega_{l,j'}} = \sum_{l=1}^L \E{ \omega_{l,j} \, \omega_{l,j'} } \overset{\mathrm{ind.}}{=} \sum_{l=1}^L \E{ \omega_{l,j} }^2 = \frac{1}{L}.
\end{equation*}

\begin{figure}[th]
    \centering
    \includegraphics[width=\linewidth]{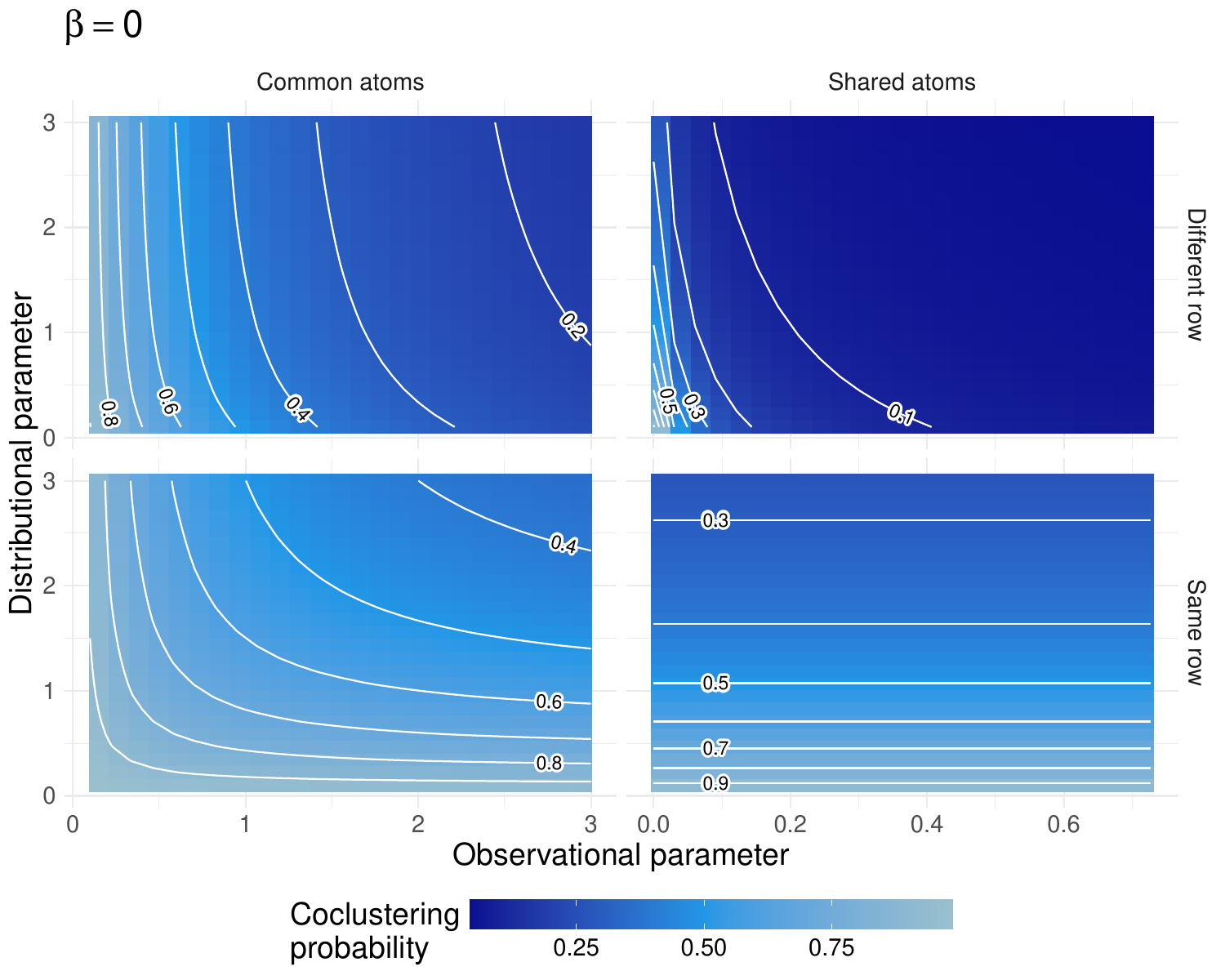}
    \caption{Heatmaps depicting the a priori coclustering probabilities for pairs of matrix entries in a two-sample scenario, plotted against the observational and distributional parameters governing the nested mixture weights. The left column illustrates the case with common atoms, while the right column shows the case with shared atoms. The top panels display probabilities for entries on different rows. In contrast, the bottom panels show the corresponding probabilities for entries on the same row. The inverse temperature parameter is fixed at $\beta = 0$.}
    \label{fig:cocl1}
\end{figure}

\begin{figure}[th]
    \centering
    \includegraphics[width=\linewidth]{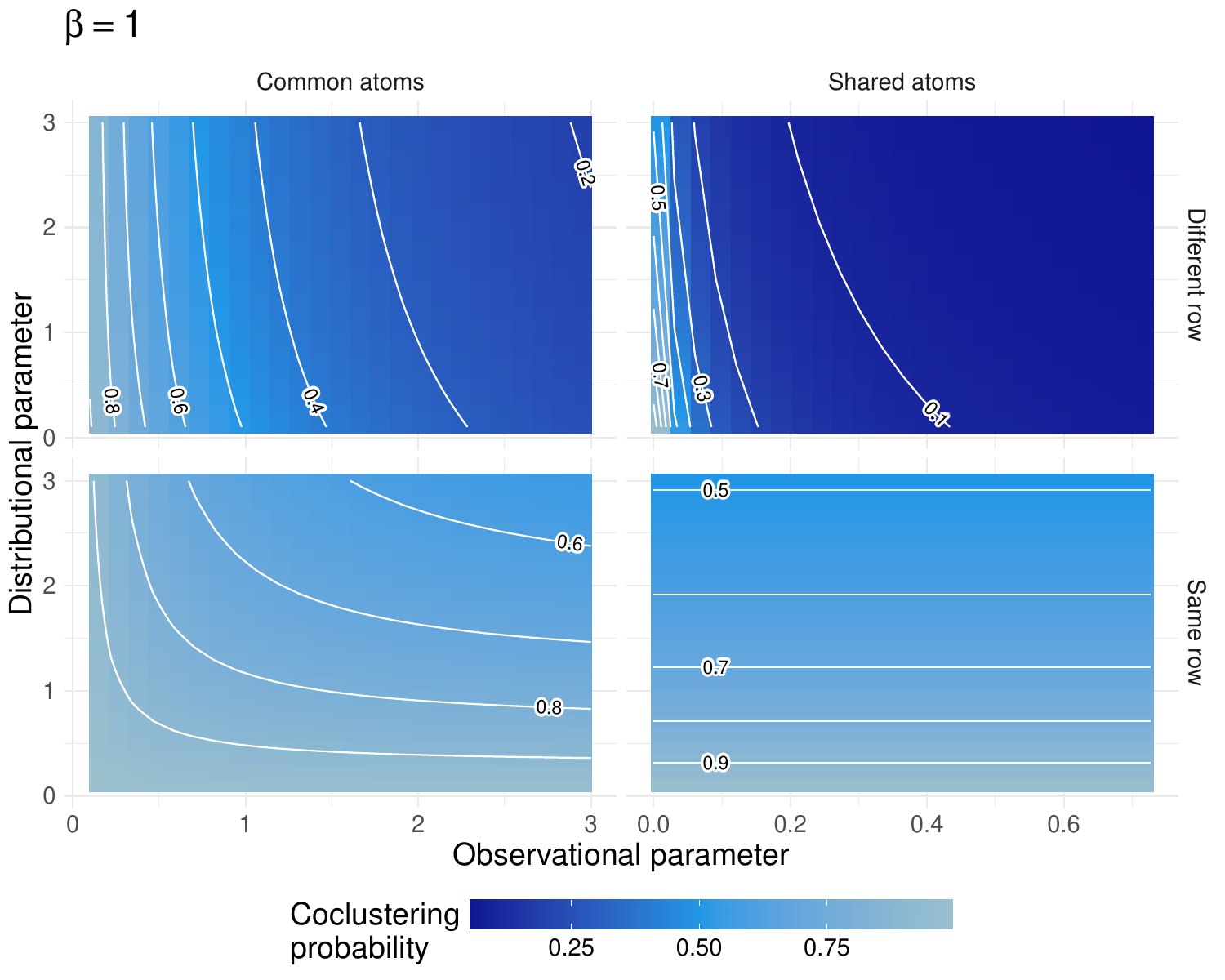}
    \caption{Heatmaps depicting the a priori coclustering probabilities for pairs of matrix entries in a two-sample scenario, plotted against the observational and distributional parameters governing the nested mixture weights. The left column illustrates the case with common atoms, while the right column shows the case with shared atoms. The top panels display probabilities for entries on different rows. In contrast, the bottom panels show the corresponding probabilities for entries on the same row. The inverse temperature parameter is fixed at $\beta = 1$.}
    \label{fig:cocl2}
\end{figure}

\begin{figure}[th]
    \centering
    \includegraphics[width=\linewidth]{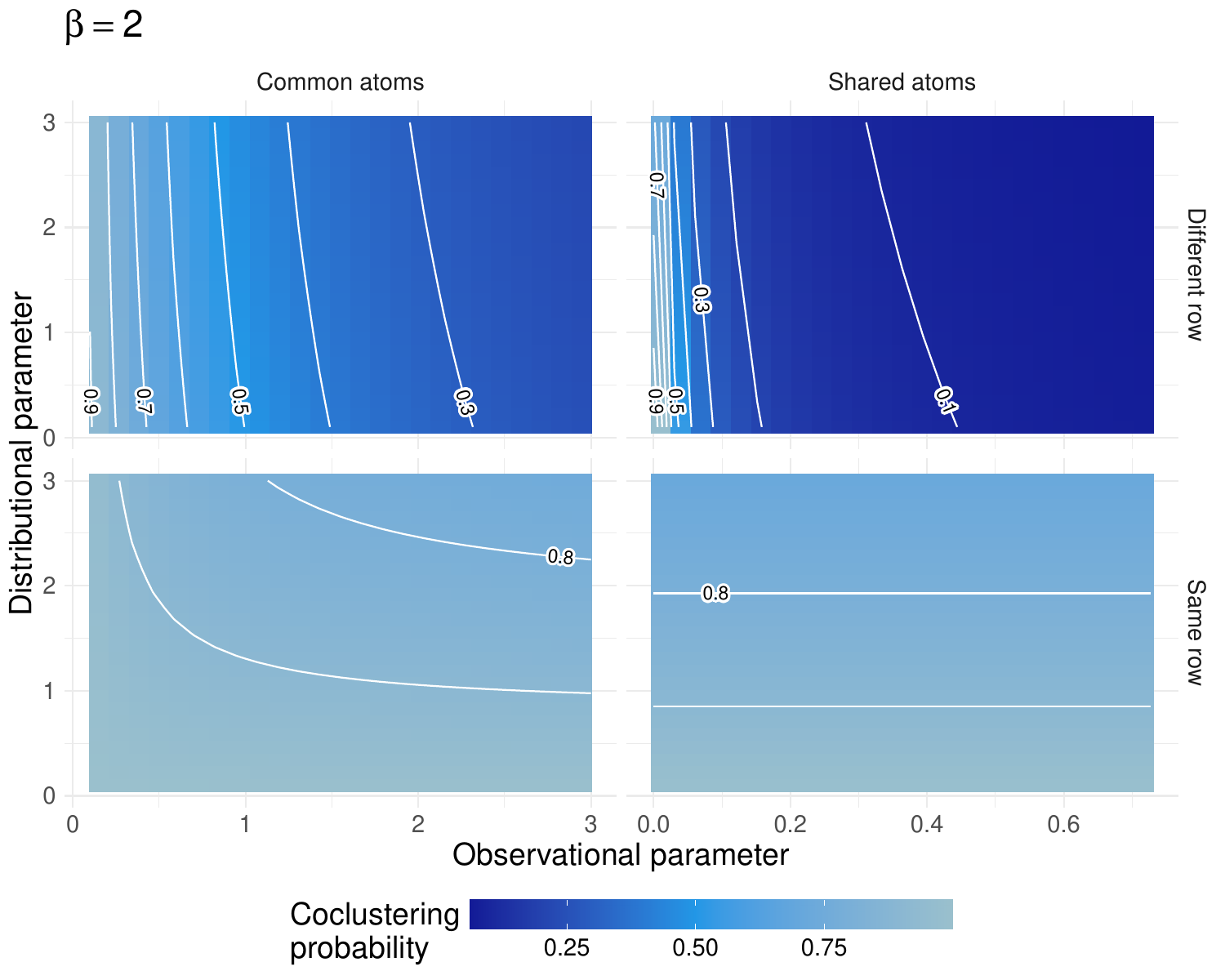}
    \caption{Heatmaps depicting the a priori coclustering probabilities for pairs of matrix entries in a two-sample scenario, plotted against the observational and distributional parameters governing the nested mixture weights. The left column illustrates the case with common atoms, while the right column shows the case with shared atoms. The top panels display probabilities for entries on different rows. In contrast, the bottom panels show the corresponding probabilities for entries on the same row. The inverse temperature parameter is fixed at $\beta = 2$.}
    \label{fig:cocl3}
\end{figure}

In Figures~\ref{fig:cocl1}--\ref{fig:cocl3}, we examine the behavior of the derived prior coclustering probabilities for two entries in the data matrix under various parameter specifications. Recall that in the common atom model, $\boldsymbol{\pi} \sim \mathrm{GEM}(\alpha)$ and $\boldsymbol{\omega}_k \sim \mathrm{GEM}(\nu)$ for all $k$, whereas in the shared atoms model, $\boldsymbol{\omega}_k \sim \mathrm{Dirichlet}_L(\boldsymbol{b}_0)$ for all $k$. Here, $\alpha$ is a \textit{distributional parameter}, while $\nu$ and $\boldsymbol{b}_0$ are \textit{observational parameters}. We compute these probabilities for three inverse temperature levels, i.e., $\beta \in \{0, 1, 2\}$. The values of $\alpha$ and $\nu$ are varied over the interval $(0, 3)$, and the Dirichlet concentration parameters are constrained to the unitary interval to enforce sparsity. We present comparisons for both the same-row and different-row scenarios.

\FloatBarrier
\clearpage
\subsection{Approximations in the CAVI algorithm}
\label{supp:estimating_beta}
Due to the presence of an intractable normalizing constant, inference on the terms involving the inverse temperature parameter of a Potts model $\beta$ is extremely challenging. Many approaches have been proposed in the literature, ranging from the use of approximations such as the \emph{pseudo-likelihood} \citep[PL,][]{Besag_lattice}, to more refined samplers \citep{Murray2006}, to ABC techniques \citep{Moores2020}. 

In the VB framework, \cite{Lu2020, Durand2022} and treat $\beta$ as a fixed nuance parameter w.r.t. maximizing the lower bound, thus resorting to a VB-Expectation Maximization (VBEM) algorithm. This approach translates into computing 
\begin{equation}
    \hat\beta = \arg\max_\beta \mathbb{E}_{ q_{\boldsymbol{C}},q_{\boldsymbol{v}}}\left[{\log(p(\boldsymbol{C}\mid \boldsymbol{\pi};\beta))}\right].
    \label{eq:maxbeta}
\end{equation}

Conversely, \citet{McGrory2009} treated $\beta$ as a random variable and proposed different approximations to the normalizing constant to derive the variational distribution $q(\beta)$. In our algorithm, we adopt approximations inspired by~\citet{McGrory2009}, which we delineate in the following. 

A key quantity for both the approximate updates of $\beta$ and the ELBO components for both $\boldsymbol{C}$ and $\beta$ is the expectation of $\log p(\boldsymbol{C}\mid \boldsymbol{\pi},\beta)$. Given the intractability of the normalizing constant involved in the expression, computing this expectation analytically is unfeasible. We then resort to two approximations. First, the Potts prior is approximated using a mean-field pseudo-likelihood (PL) approach. Then, every variable in $\boldsymbol{v}$ is substituted with its variational expectation (A2), switching $\log$ and expectation. Therefore, we first write
\begin{equation}
    p(\boldsymbol{C}\mid \boldsymbol{\pi},\beta)  \stackrel{(PL)}{\approx}\prod_{j=1}^J\tilde p(C_j\mid \boldsymbol{C}_{-j},\boldsymbol{\pi},\beta) = \prod_{j=1}^J \frac{\exp\left[\log\pi_{C_j} +\beta \sum_{i \in \partial_j}\mathds{1}_{\{C_i=C_j\}}\right]}{\sum_{k\geq 1}\exp\left[\log\pi_k +\beta \sum_{i \in \partial_j}\mathds{1}_{\{C_i=k\}}\right]},
\end{equation}

Then, taking the expectation w.r.t. all the variables but $\beta$, we have
\begin{align*}
         \E{\log p(\boldsymbol{C}\mid \boldsymbol{\pi},\beta)}  &\stackrel{(PL)}{\approx}\sum_{j=1}^J\E{\sum_{k\geq 1}\mathds{1}_{\{C_j=k\}}\log\pi_k + \beta \sum_{k\geq 1}\sum_{i \in \partial_j} \mathds{1}_{\{C_i=k\}}\mathds{1}_{\{C_j=k\}} }\\&-\E{\sum_{j=1}^J\log \sum_{k\geq 1}\exp\left[\log\pi_k +\beta \sum_{i \in \partial_j}\mathds{1}_{\{C_i=k\}}\right]}\\
     &= \sum_{j=1}^J\sum_{k\geq 1} \rho_{j,k}\E{\log\pi_k} + 
     \beta \sum_{j=1}^J\sum_{k\geq 1}\sum_{i \in \partial_j}\rho_{i,k}\rho_{j,k}\\&-\sum_{j=1}^J\E{\log \sum_{k\geq 1}\exp\left[\log\pi_k +\beta \sum_{i \in \partial_j}\mathds{1}_{\{C_i=k\}}\right]}.\\
\end{align*}

We can further approximate this term by writing

\begin{align*}
     &\stackrel{(A2)}{\approx}
     \sum_{j=1}^J\sum_{k\geq 1} \rho_{j,k}\left[ \E{\log\pi_k} + 
     \beta \sum_{i \in \partial_j}\rho_{i,k}\right]-\sum_{j=1}^J \log \sum_{k\geq 1} \exp\left[\E{\log\pi_k} +\beta \sum_{i \in \partial_j} \rho_{j,k}\right],
\end{align*}
where we set ${\rho}_{j,k}=q(C_j=k)$. The two approximations led to a quantity that is easy to evaluate. Thus, given a prior $p(\beta)$ for the inverse temperature parameter, we can approximate
\begin{equation}
    q^*(\beta) \stackrel{(PL+A2)}{\propto} p(\beta) \prod_{j=1}^J\frac{ \exp\left[\sum_{k\geq 1} \rho_{j,k}\left[ \E{\log\pi_k} + 
     \beta \sum_{i \in \partial_j}\rho_{i,k}\right]\right]}{
     \sum_{k\geq 1} \exp\left[\E{\log\pi_k} +\beta \sum_{i \in \partial_j} \rho_{j,k}\right]}.
\end{equation}

Suppose we assume $p(\beta)$ to be a uniform distribution over a grid of sensible values. In that case, it is then straightforward to evaluate the variational expected value of this distribution.

\subsection{Poseidon: Elbo components}
\label{supp:elboposeid}
\begin{itemize}
\item     $ \mathbb{E}\left[\log p(\boldsymbol{Y} \mid \boldsymbol{R},\boldsymbol{C}, \btheta)\right] = \sum_{t=1}^{T}\sum_{j=1}^J \sum_{i=1}^{N} \sum_{k=1}^K \sum_{l=1}^L \rho_{j,k}\xi^{(t)}_{i,k,l}\ell^{(t)}_{i,j,l}$   
    where\\$\ell^{(t)}_{i,j,l} =\mathbb{E}\left[\log(\phi(y^{(t)}_{i,j}|\theta^{(t)}_l))\right] =  -\frac12 \left[ \log(d^{(t)}_l) - \psi(c^{(t)}_l) +\frac{1}{k^{(t)}_l}+\frac{c^{(t)}_l}{d^{(t)}_l} (y^{(t)}_{i,j}-m^{(t)}_l)^2\right]$ ;

    \item $\mathbb{E}\left[\log p(\boldsymbol{R} \mid\bomega)\right] =
    \sum_{t=1}^T \sum_{i=1}^{N} \sum_{k=1}^K \sum_{l=1}^L   \: \xi^{(t)}_{i,k,l}  \: h_l\left(\boldsymbol{p}^{(t)}_k\right); $
    
    \item $\mathbb{E}\left[\log q(\boldsymbol{R})\right] = \sum_{t=1}^{T}\sum_{i=1}^{N}\sum_{k=1}^K \sum_{l=1}^L \xi^{(t)}_{i,k,l}\log(\xi^{(t)}_{i,k,l})$;

    \item $\mathbb{E}\left[\log p(\boldsymbol{C}\mid \boldsymbol{\pi},\beta)\right]$ is not available in closed form. One can approximate it, following the rationale presented in Section~\ref{supp:estimating_beta} as 
    \begin{align*}
    \mathbb{E}\left[\log p(\boldsymbol{C}\mid \boldsymbol{\pi},\beta)\right] &\approx   \sum_{j=1}^J\sum_{k\geq 1} \rho_{j,k}\left[ \E{\log\pi_k} + 
     \bar\beta \sum_{i \in \partial_j}\rho_{i,k}\right]\\&-\sum_{j=1}^J \log \sum_{k\geq 1} \exp\left[\E{\log\pi_k} +\bar\beta \sum_{i \in \partial_j} \rho_{j,k}\right],
     \end{align*}
    with $\bar\beta$ representing the expected value of the variational distribution for $\beta$, computed numerically.
    
    \item $\mathbb{E}\left[\log q(\boldsymbol{C})\right] = \sum_{j=1}^{K}\sum_{k=1}^K  \rho_{j,k}\log( \rho_{j,k})$

    \item $\mathbb{E}\left[\log p(\boldsymbol{v}\mid \alpha)\right] =    (K-1)\left(\psi(s_1)-\log(s_2)\right) + \left(s_1/s_2-1\right)
    \sum_{k=1}^{K-1} g(\bar{b}_{k},\bar{a}_{k}); $

    \item $\mathbb{E}\left[\log q(\boldsymbol{v})\right] = 
\sum_{k=1}^K  \{\log(\mathcal{C}_{\boldsymbol{v}}(\bar{a}_{k},\bar{b}_{k}))+
    (\bar{a}_{k}-1)g(\bar{a}_{k},\bar{b}_{k})+
   (\bar{b}_{k}-1)g(\bar{b}_{k},\bar{a}_{k})\}$, where $\mathcal{C}_{\bv}(\cdot)$ is the normalizing constant of a Beta distribution;
    
    \item $\mathbb{E}\left[\log p(\boldsymbol{\omega})\right] =
    T\sum_{k=1}^K \log(\mathcal{K}_{\boldsymbol{\omega}_k}(\boldsymbol{b}_0))+
     \sum_{t=1}^T\sum_{k=1}^K \sum_{l=1}^L (b^{(t)}_{l,k}-1) h^{(t)}_l(\boldsymbol{p}^{(t)}_k),$
    where $\mathcal{K}_{\boldsymbol{\omega}}(\cdot)$ is the normalizing constant of a $\Dirichlet_L$ distribution;

    \item $\mathbb{E}\left[\log q(\bomega)\right] = \sum_{t=1}^T\left[\sum_{k=1}^K \log(\mathcal{K}_{\bomega}(\boldsymbol{p}^{(t)}_k))+  \sum_{k=1}^K \sum_{l=1}^L (p^{(t)}_{l,k}-1) h_l(\boldsymbol{p}^{(t)}_k)\right]$;

        \item $\mathbb{E}\left[\log p(\boldsymbol{\mu},\boldsymbol{\sigma}^2)\right] =
    TL\log(\mathcal{K}_{(\mu,\sigma^2)}(m_0,\kappa_0,\tau_0,\gamma_0)) - (\tau_0+1+1/2)\sum_{t=1}^T\sum_{l=1}^L (\log(d^{(t)}_l)-\psi(c^{(t)}_l)) - \gamma_0\sum_{t=1}^T\sum_{l=1}^L c^{(t)}_l/d^{(t)}_l-
    \frac{\kappa_0}{2}\sum_{t=1}^T\sum_{l=1}^L (\frac{1}{k^{(t)}_l} + \frac{c^{(t)}_l}{d^{(t)}_l}(m_l-m_0)^2),$        
    where $\mathcal{K}_{(\mu,\sigma^2)}(\cdot)$ is the normalizing constant of a Normal-Inverse Gamma distribution;
    \item $\mathbb{E}\left[\log q(\boldsymbol{\mu},\boldsymbol{\sigma}^2)\right] = \sum_{t=1}^T\sum_{l=1}^L \log(\mathcal{K}_{(\mu^{(t)},\sigma^{2,(t)})}(m^{(t)}_l,k^{(t)}_l,c^{(t)}_l,d^{(t)}_l)) - \sum_{t=1}^T\sum_{l=1}^L  (c^{(t)}_l +1+ 1/2)(\log(d^{(t)}_l) - \psi(c^{(t)}_l)) - \sum_{t=1}^T\sum_{l=1}^L  c^{(t)}_l - (L\cdot T)/2$;

    \item 
    $\mathbb{E}\left[\log p(\alpha)\right] =  \log(\mathcal{K}_{\alpha}(a_\alpha,b_\alpha)) + (a_\alpha-1)(\psi(s_1)-\log(s_2)) - b_\alpha s_1/s_2$, where $\mathcal{K}_{\alpha}(\cdot)$ is the normalizing constant of a Gamma distribution.

    \item $\mathbb{E}\left[\log q(\alpha)\right] = \log(\mathcal{K}_{\alpha}(s_1,s_2)) + (s_1-1)(\psi(s_1)-\log(s_2)) - s_1$.

    \item If $\beta$ is random and uniformly distributed over a grid, the difference in the ELBO $\mathbb{E}\left[\log p(\beta)-\log q(\beta)\right]$ can be approximated with $\mathbb{E}\left[\log p(\boldsymbol{C})\right]$, which then simplifies with the ELBO update given by the variational distribution of the CC membership labels $\boldsymbol{C}$.

\end{itemize}

\clearpage
\section{Additional details about the simulation studies}

\subsection{Biclustering recovery simulation study}
\label{supp_sec_biclrecovery}
In Section~\ref{sec:simulation1} of the manuscript, we present a simulation study to analyze the biclustering recovery performance of Pose, compared to other biclustering methods. We now provide additional details regarding the synthetic data generation and hyperparameter choices when running the different algorithms.

\begin{figure}[t!]
    \centering
    \includegraphics[width=0.5\linewidth]{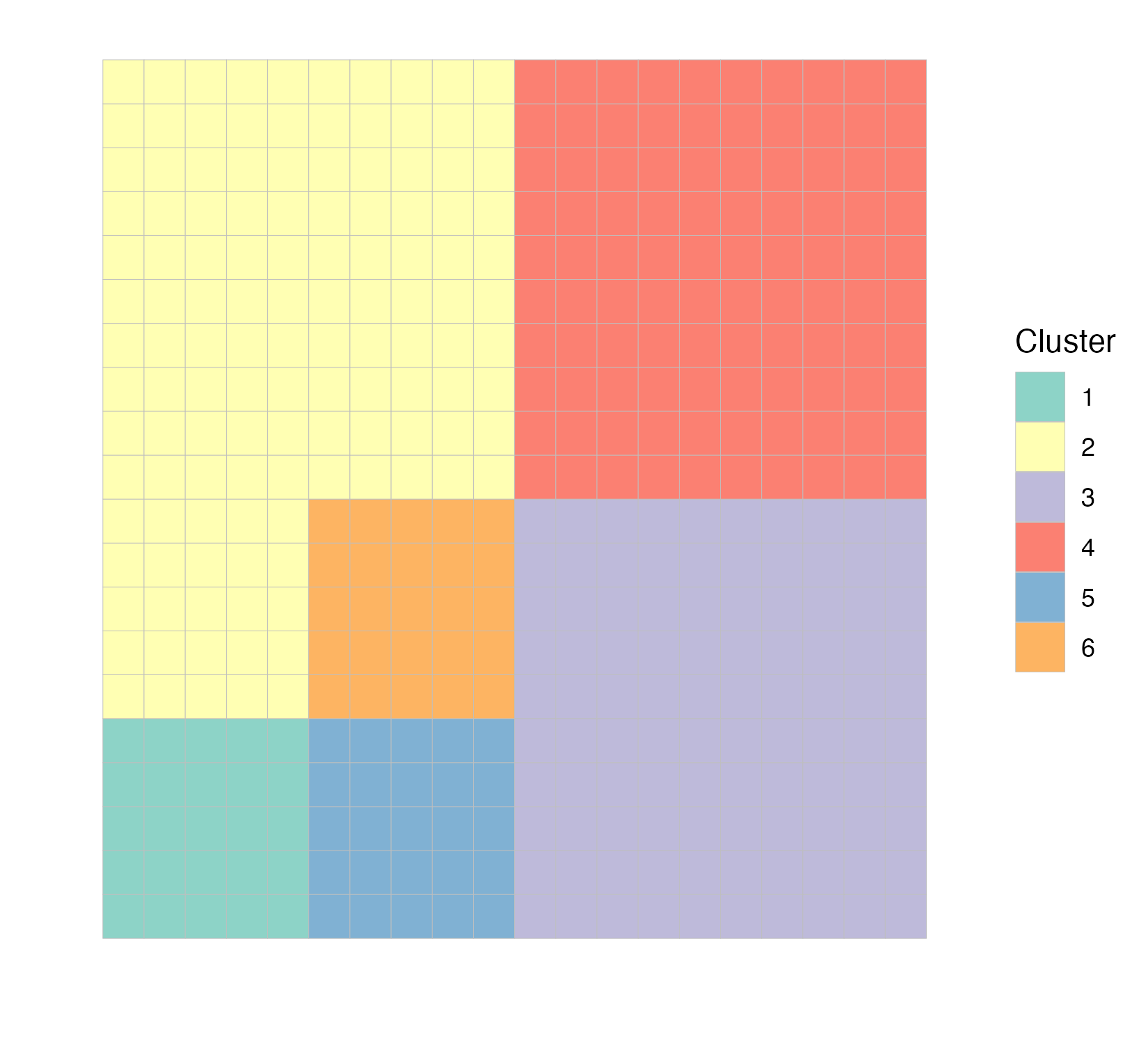} 
    \caption{The column partition structure used for the simulation study described in Section~\ref{sec:simulation1} of the manuscript.}
    \label{fig:column_cluster_simB}
\end{figure}

\begin{figure}[t!]
    \centering
    \includegraphics[width=0.44\linewidth]{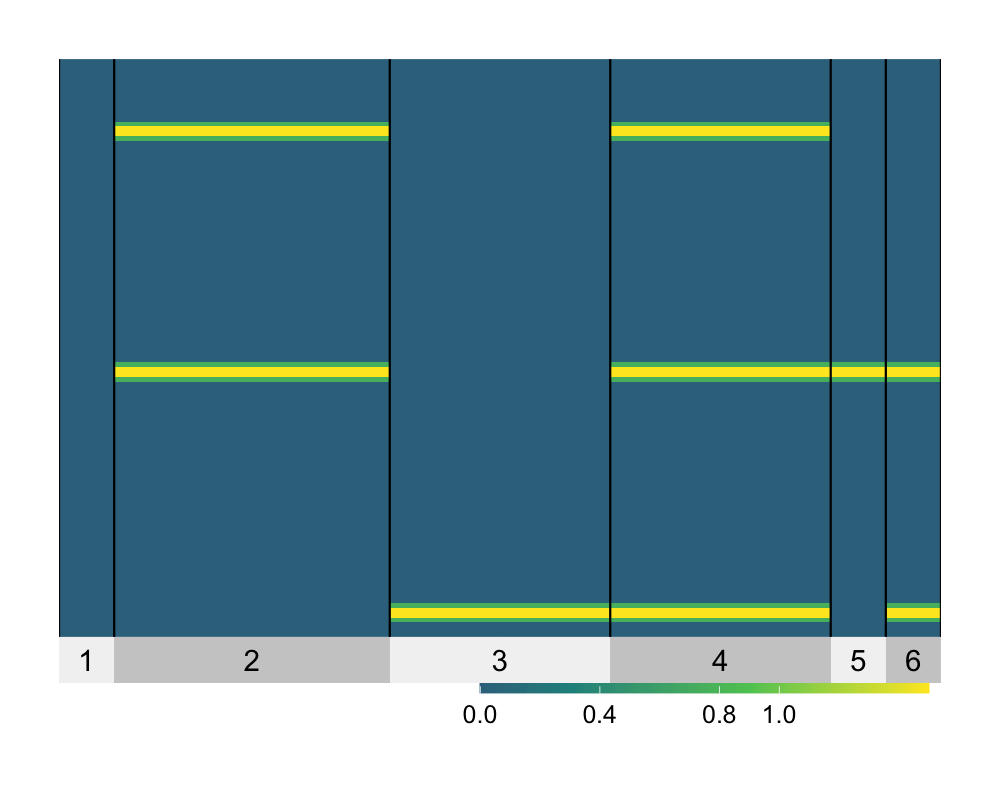} ~
    \includegraphics[width=0.53\linewidth]{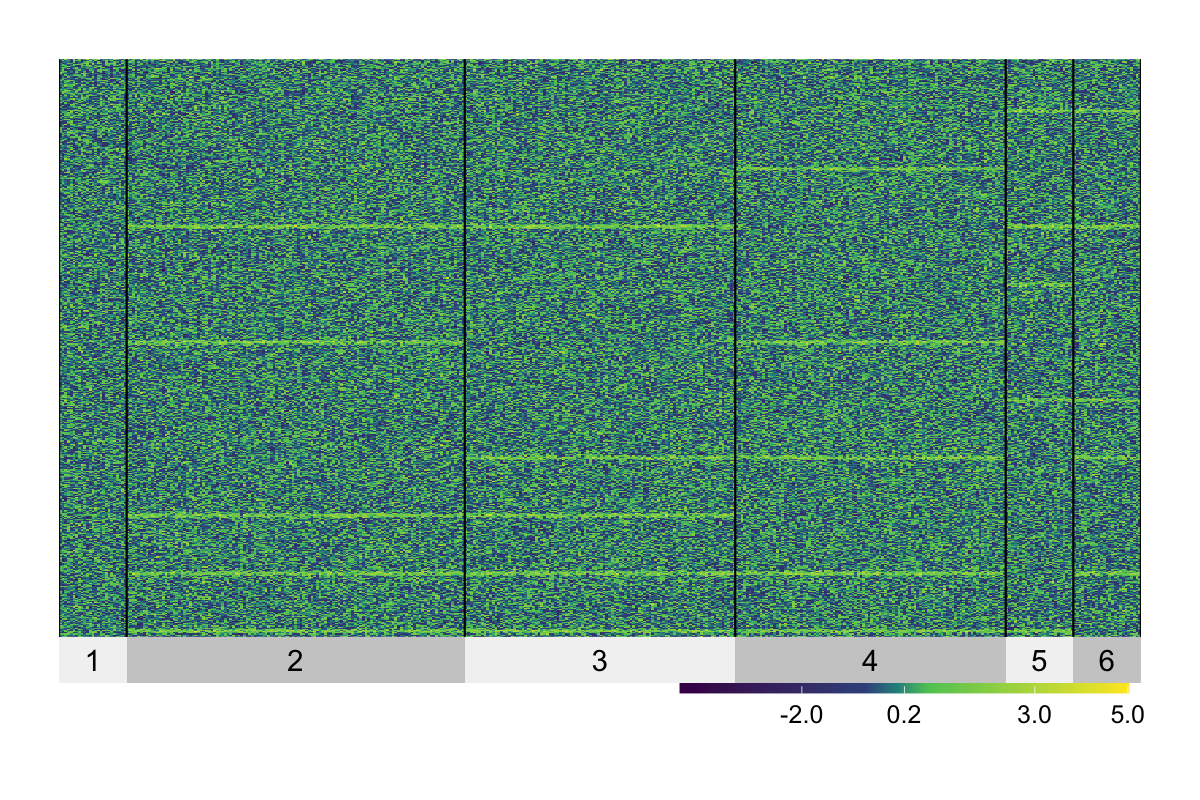}
    \caption{\textit{Left}: an example of the biclustering structure (with row cluster allocations within each column cluster), for a subset of 120 rows and all columns (ordered and labeled by their column clusters). The colors represent the mean activation for each element. \textit{Right}: an example of the data generated given the biclustering allocation (for all rows and columns). Colors correspond to the intensity activation level of each \emph{m/z} in the corresponding pixel, and columns have been sorted by their column clusters.}
    \label{fig:data_simB}
\end{figure}

The synthetic datasets are generated to mimic some characteristics observed in MALDI-MSI data. Specifically, we consider a grid of $20 \times 20$ pixels, with a fixed partition structure (shared by all synthetic datasets) that divides the pixels into six clusters, each containing 25, 25, 25, 100, 100, and 125 contiguous pixels, respectively. A visual representation of the partition structure is provided in Figure~\ref{fig:column_cluster_simB}. 
We consider 500 rows, each corresponding to a hypothetical \emph{m/z} level. To generate the \emph{m/z} intensity level and create a biclustering structure, for each column cluster, we assign a number of rows to a group of ``high-activation'' \emph{m/z} values and other rows to a group of ``medium-activation'' \emph{m/z} values. The remaining rows are classified as ``noise''. These three categories correspond to the row clusters that are generated for each column cluster.
The rows assigned to the ``high-activation'' \emph{m/z}, which correspond to the activation peaks, are selected randomly, so that there are five groups of peaks, each corresponding to 2 adjacent rows. Note that different column clusters display different sets of ``high-activation'' \emph{m/z}, and by default, the first column cluster is designed to have no ``high-activation'' \emph{m/z}. Given the group of rows assigned to the ``high-activation'' \emph{m/z}, the ``medium-activation'' \emph{m/z} are selected by considering the rows immediately above and below the ``high-activation'' \emph{m/z}. 
Given each row's allocation to these groups, we generate the activation levels from a Gaussian distribution with means of 1.5, 0.75, and 0 for the high-activation, medium-activation, and noise groups, respectively, and a standard deviation of 1. 
A visual example of the row clustering allocation is provided in the left panel of Figure~\ref{fig:data_simB}, where only a subset of 120 rows is displayed for visualization purposes, and the matrix entries have been reordered by the column cluster assignment. The colors correspond to the mean intensity associated with each \emph{m/z} (row) in the corresponding pixel (column).
The right panel of Figure~\ref{fig:data_simB} instead visualizes the entire data matrix for one instance of the data generation process. Colors represent the activation level of each \emph{m/z} in the corresponding pixel.

Given the spatial structure of the chosen column clustering partition (see Figure~\ref{fig:column_cluster_simB}), we fit the Pose model with a strong spatial prior fixing $\beta = 1$. We set the upper bound on the number of clusters as $L = 10$ and $K = 10$. The hyperparameters are set across all $l$ as follows: $(m_0, k_0, c_0, d_0) = \left(0, 0.1, 3, 2\right)$, $\boldsymbol{b}_0 = 0.0001 \cdot \boldsymbol{1}_L$, and $s_1 = s_2 = 1$. We run the CAVI algorithm from 10 different starting points and retain only the one that reaches the highest ELBO value. This process took, on average, 12 seconds for each dataset on a MacBook Pro with an 8-core Apple M1 Pro processor and 32GB unified memory.

The other biclustering models we consider do not have a spatial component, but can still be fit to spatial data. The \texttt{sparseBC} algorithm \citep{tan2014sparse} was run both fixing the number of column clusters $K$ to the true value 6 (\texttt{trueK}) and using their function \texttt{sparseBC.chooseKR} that performs cross-validation to select the number of column and row clusters (\texttt{chooseKR}). In both cases, we set the regularization parameter $\lambda$ to $10$. 
The \texttt{Double k-means} algorithm was designed to mimic the main ideas behind the Pose model. First, the k-means algorithm is run on the columns of the matrix, and the number of clusters is chosen using the gap method (using the \texttt{cluster::clusGap} function in R \citep{tibshirani2001estimating,cluster_package}. Given the selected column clusters, the average intensity is computed for each \emph{m/z} across all columns belonging to the same column cluster. Another k-means algorithm is run on these average intensity values, determining the row cluster allocation for each column cluster. The number of clusters is again selected using the gap method. 

To compute the Adjusted Rand Index (ARI) between the true and estimated biclustering structures, the biclustering labels are vectorized, transforming the cluster allocation matrix into vectors for both the true and estimated allocations. Then, the ARI is applied between these vectors. 
To evaluate the biclustering RMSE, when the mean values associated with each cluster are not estimated by the algorithm, we compute the empirical average of the activation levels in each estimated bicluster and compare this with the generated mean activation levels (ground truth) to calculate the RMSE. 

\subsection{Informative datasets guide the shared clustering estimation}
\label{supp:sec_simu_D0D4}

In Section~\ref{sec:simulation2} of the manuscript, we present a simulation study to analyze the improvement of the partition estimation of Poseidon when compared to a simple, one-dataset Pose model. Here, we now describe in more detail the synthetic data generation and then present the hyperparameter choices we made.

We consider five datasets, denoted as ${D}^{(r)}$, each comprising $J=20$ columns and $N^{(r)}=50$ rows, with $r = 0, 1, 2, 3, 4$. A common biclustering structure is assumed across these datasets. As ground truth, we assume the presence of two column clusters, each consisting of 10 columns. Specifically, the first column cluster contains 25 rows where cell entries are sampled from a Normal distribution with mean $\mu^{(r)}$ and variance 1. The remaining 25 rows in this cluster are generated from a standard Gaussian distribution. The second column cluster contains a single RC, where all cell entries are sampled from a standard Gaussian distribution. 

First, we generate a reference dataset, ${D}^{(0)}$, with $\mu^{(0)} = 4$. This case represents a scenario where a clear separation exists among the row clusters and, consequently, among the column clusters. For the remaining datasets, ${D}^{(r)}$ with $r = 1, 2, 3, 4$, we set $\mu^{(r)} \in \{1, 0.75, 0.50, 0.25\}$. A visual example of the realization of these matrices is reported in the Supplementary Figure~\ref{suppfig:datasim}. In these datasets, the overlap between the distributions characterizing the row clusters increases as $r$ increases, with the noise increasingly dominating the signal, making it harder to distinguish the underlying true partition. A visual example of these datasets is reported in Figure~\ref{suppfig:datasim}.
\begin{figure}[ht]
    \centering
    \includegraphics[width=1\linewidth]{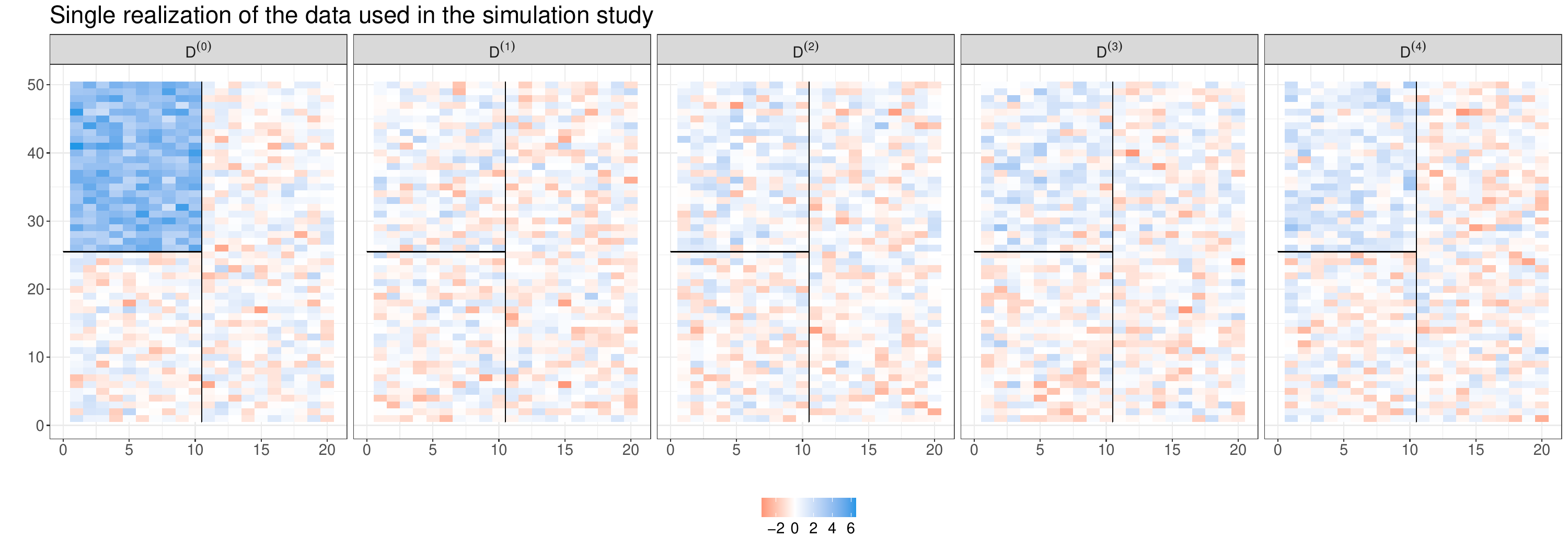}
    \caption{Example of realization of a collection of matrix datasets used in the second simulation study.}
    \label{suppfig:datasim}
\end{figure}

We focus on the Pose and Poseidon models without spatial interaction by setting $\beta = 0$. The hyperparameters are set, for $t=1,\ldots,T$ as $(m^{(t)}_0, k^{(t)}_0, c^{(t)}_0, d^{(t)}_0) = \left(0, 0.1, 3, 2\right)$, $\boldsymbol{b}_0 = 0.0001 \cdot \boldsymbol{1}_L$, and $s_1 = s_2 = 1$. Additionally, we set $L = 30$ and $K = 15$, and we consider the best CAVI run out of 50 random initializations. Each experiment is simulated 100 times to examine the Monte Carlo variability of the results.

As discussed in the manuscript, our findings reveal two key insights. First, the inclusion of the reference dataset improves the recovery of shared column partition, effectively compensating for noise present in the other datasets. Second, the multi-dataset framework significantly enhances the recovery of row clusters in noisy datasets compared to single-dataset analyses, while maintaining the same quality for the reference dataset. 

\begin{figure}[t!]
    \centering
    \includegraphics[width=\linewidth]{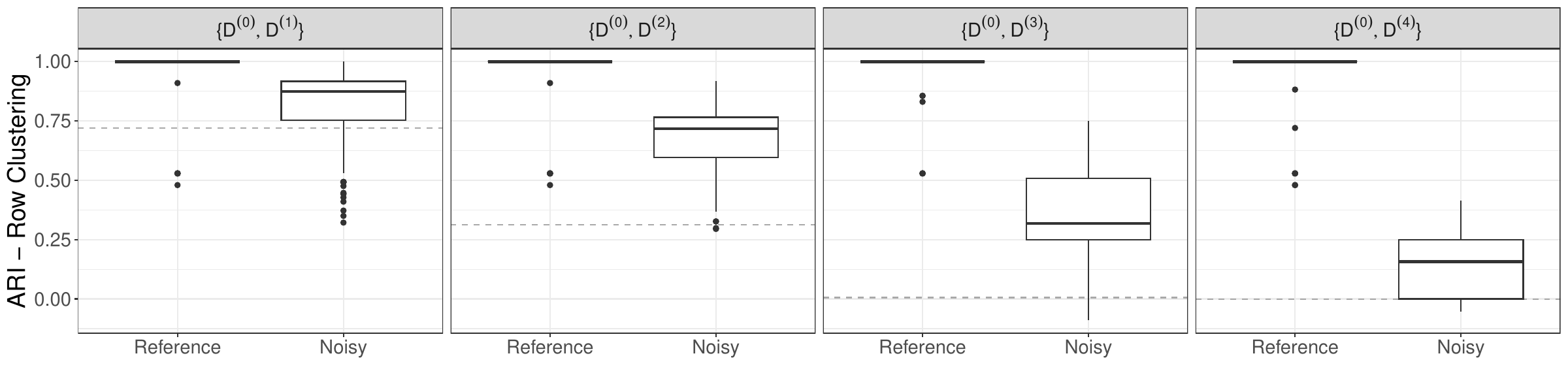}
    \caption{Boxplots of the ARI values for the row clustering solutions obtained across the 100 replicas of the experiment. Each panel shows the reference and noisy data results, stratified by dataset collection. The horizontal dashed lines display the average ARI obtained with the noisy datasets in the single-dataset modeling framework.}
    \label{suppfig:RC_ARI_IDON}
\end{figure}

To complement the results in Table~\ref{tab:DCARI} of the main manuscript, Figure~\ref{suppfig:RC_ARI_IDON} presents the boxplots reporting the distributions of the ARI values estimated for the row clusters across the different dataset collections analyzed with Poseidon. In addition, Table~\ref{supp:tab:NCLUST_RC} reports the average and standard deviation of the estimated number of row clusters for three settings: the single-dataset case (top row), Poseidon applied to the multi-dataset collections $\{D^{(0)}, D^{(r)}\}$ (central rows), and Pose applied to the stacked datasets $\left[D^{(0)}, D^{(r)}\right]$ (bottom row), for $r = 1, \ldots, 4$.

\begin{table}[t!]
    \centering
    \begin{tabular}{lcccccccccc}
    \toprule
     Pose & \multicolumn{2}{c}{${D}^{(1)}$} & \multicolumn{2}{c}{${D}^{(2)}$} & \multicolumn{2}{c}{${D}^{(3)}$} & \multicolumn{2}{c}{${D}^{(4)}$}\\
    \cmidrule(lr){1-1} \cmidrule(lr){2-3} \cmidrule(lr){4-5} \cmidrule(lr){6-7} \cmidrule(lr){8-9} 
    RC &  2.37 & (0.485) & 1.69 & (0.598) & 1.06 & (0.239) & 1.03 & (0.171)\\
    \midrule
    Poseidon & \multicolumn{2}{c}{$\{D^{(0)},D^{(1)}\}$} & \multicolumn{2}{c}{$\{D^{(0)},D^{(2)}\}$} & \multicolumn{2}{c}{$\{D^{(0)},D^{(3)}\}$} & \multicolumn{2}{c}{$\{D^{(0)},D^{(4)}\}$}\\
    \cmidrule(lr){1-1} \cmidrule(lr){2-3} \cmidrule(lr){4-5} \cmidrule(lr){6-7} \cmidrule(lr){8-9} 
   RC Ref. & 2.07 & (0.256) & 2.08 & (0.273) & 2.07 & (0.256) & 2.09 & (0.321)\\
   RC Noisy & 2.18 & (0.386) & 2.15 & (0.359) & 2.09 & (0.404) & 1.66 & (0.572)\\
        \midrule
      Stacked & \multicolumn{2}{c}{$[D^{(0)},D^{(1)}]$} & \multicolumn{2}{c}{$[D^{(0)},D^{(2)}]$} & \multicolumn{2}{c}{$[D^{(0)},D^{(3)}]$} & \multicolumn{2}{c}{$[D^{(0)},D^{(4)}]$}\\
    \cmidrule(lr){1-1} \cmidrule(lr){2-3} \cmidrule(lr){4-5} \cmidrule(lr){6-7} \cmidrule(lr){8-9} 
    RC & 3.29 & (0.498) & 3.33 & (0.514) & 3.06 & (0.422) & 2.62 & (0.508)\\
    \bottomrule
    \end{tabular}
    \caption{Average and standard deviation of the number of unique row clusters (RC) using Pose, Poseidon, and stacking the reference and noisy datasets. The number of RCs estimated with Pose on the reference dataset $D^{(0)}$ is 2.22 (0.416).}
    \label{supp:tab:NCLUST_RC}
\end{table}

\clearpage
\section{Additional results of Poseidon on ccRCC tissue}
\label{supp:other_mol}
\paragraph{Additional insights about \emph{m/z}}
In the main paper, we demonstrated how the image segmentation estimated by Poseidon effectively captures distinct multiomics clusters, reflecting the complex histopathological landscape of the kidney tumor resection previously reported by \citet{denti2022spatial}. This segmentation mirrors known tissue heterogeneity and provides a framework for identifying region-specific molecular signatures. Pinpointing the most influential molecules in each area could offer valuable insights for biomedical research, particularly in uncovering potential diagnostic markers and therapeutic targets. As a proof of concept, in Section~\ref{sec:application}, we focused on three representative clusters: a benign region (central row of Figure~\ref{fig:fig5_three_clusters}, Cluster B), a tumor area adjacent to the capsule (top row, Cluster A), and a tumor region with hemorrhagic features (bottom row, Cluster C). As previously discussed, Poseidon enables us to summarize and quantify the contribution of each analyte across the three molecular classes, studying the row partitions within each CC. These results confirm that lipids, N-glycans, and peptides convey complementary biological information. 

\begin{figure*}[h]
    \centering
    \includegraphics[width=1\linewidth]{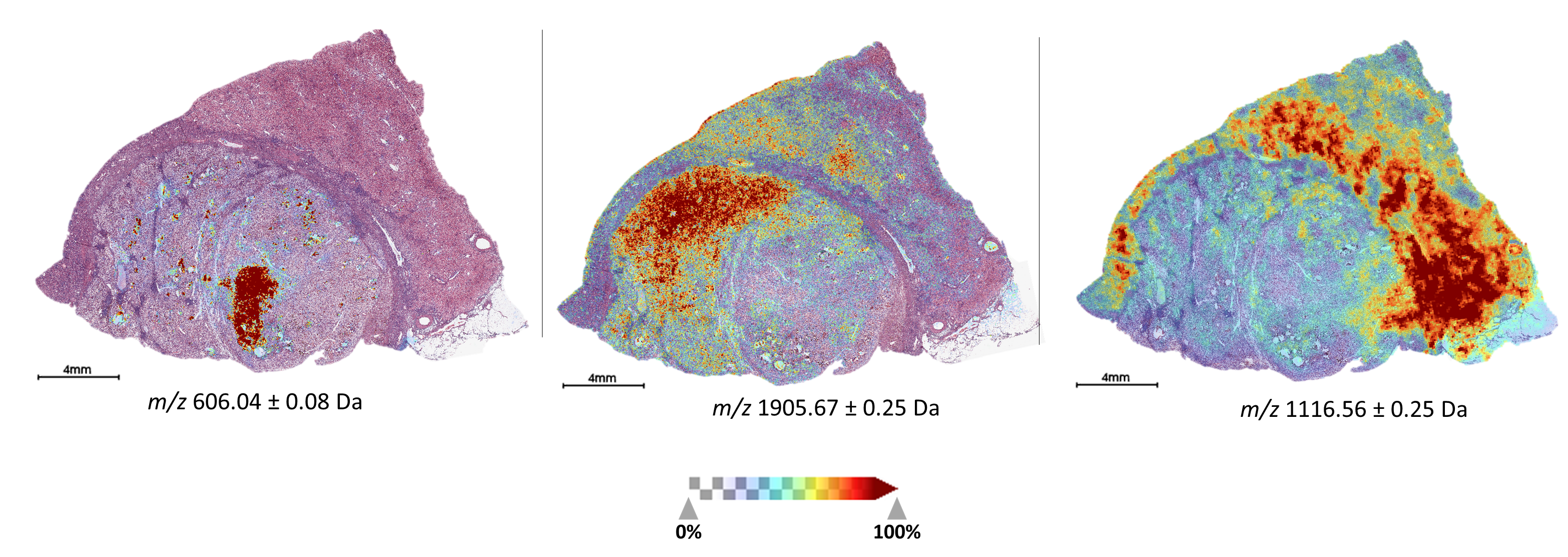}
    \label{fig:kidney}
    \caption{Spatial distribution of three selected \textit{m/z} features across kidney tissue in the three molecular classes. From left to right: lipid (\textit{m/z} 606.06), N-glycan (\textit{m/z} 1905.67), and peptide (\textit{m/z} 1116.56).}
\end{figure*}

In contrast to approaches that analyze each molecular class independently, our method integrates all modalities simultaneously, enabling direct comparison of signal intensities across distinct molecular families.
For instance, in Figure~\ref{fig:kidney}, we illustrate the spatial distribution of three representative molecules associated with specific tissue regions. In particular, given the summary provided in Figure~\ref{fig:fig5_three_clusters}, we were able to focus on a specific lipid with \emph{m/z} 606.06, which marks the hemorrhagic area; a specific N-glycan with \emph{m/z} 1905.67 (putatively identified as \texttt{Hex9HexNAc2}), which highlights the high-grade tumor (\textit{grade 2} and \textit{grade 3}) within the tumor nodule; and a specific peptide with \emph{m/z} 1116.56 (putatively identified as \texttt{Heterochromatin Protein 1-Binding Protein 3}), which is localized in the cortex surrounding the tumor nodule.

\clearpage

%% file: 03_ARXIV.bbl
\begin{thebibliography}{46}
\providecommand{\natexlab}[1]{#1}
\providecommand{\url}[1]{\texttt{#1}}
\expandafter\ifx\csname urlstyle\endcsname\relax
  \providecommand{\doi}[1]{doi: #1}\else
  \providecommand{\doi}{doi: \begingroup \urlstyle{rm}\Url}\fi

\bibitem[Alexandrov and Kobarg(2011)]{Alexandrov2011}
Theodore Alexandrov and Jan~Hendrik Kobarg.
\newblock Efficient spatial segmentation of large imaging mass spectrometry
  datasets with spatially aware clustering.
\newblock \emph{Bioinformatics}, 27\penalty0 (13):\penalty0 i230–i238, June
  2011.
\newblock ISSN 1367-4803.
\newblock \doi{10.1093/bioinformatics/btr246}.
\newblock URL \url{http://dx.doi.org/10.1093/bioinformatics/btr246}.

\bibitem[Balluff et~al.(2019)Balluff, Buck, Martin-Lorenzo, Dewez, Langer,
  McDonnell, Walch, and Heeren]{balluff2019integrative}
Benjamin Balluff, Achim Buck, Marta Martin-Lorenzo, Fr{\'e}d{\'e}ric Dewez,
  Rupert Langer, Liam~A McDonnell, Axel Walch, and Ron~MA Heeren.
\newblock Integrative clustering in mass spectrometry imaging for enhanced
  patient stratification.
\newblock \emph{PROTEOMICS--Clinical Applications}, 13\penalty0 (1):\penalty0
  1800137, 2019.

\bibitem[Balocchi et~al.(2021)Balocchi, George, and Jensen]{Balocchi2022}
Cecilia Balocchi, Edward~I. George, and Shane~T. Jensen.
\newblock {Clustering Areal Units at Multiple Levels of Resolution to Model
  Crime in Philadelphia}.
\newblock \emph{{ArXiv preprint}}, 2021.
\newblock URL \url{http://arxiv.org/abs/2112.02059}.

\bibitem[Beraha et~al.(2021)Beraha, Guglielmi, and Quintana]{Beraha2021}
Mario Beraha, Alessandra Guglielmi, and Fernando~A. Quintana.
\newblock {The Semi-Hierarchical Dirichlet Process and Its Application to
  Clustering Homogeneous Distributions}.
\newblock \emph{Bayesian Analysis}, 16\penalty0 (4):\penalty0 1187--1219, 2021.
\newblock ISSN 19316690.
\newblock \doi{10.1214/21-BA1278}.

\bibitem[Besag(1974)]{Besag_lattice}
Julian Besag.
\newblock Spatial interaction and the statistical analysis of lattice systems.
\newblock \emph{Journal of the Royal Statistical Society: Series B},
  36\penalty0 (2):\penalty0 192--236, 1974.

\bibitem[Besag(1986)]{Julian1986}
Julian Besag.
\newblock {On the Statistical Analysis of Dirty Pictures}.
\newblock \emph{Journal of the Royal Statistical Society. Series B
  (Methodological)}, 48\penalty0 (3):\penalty0 259--302, 1986.

\bibitem[Bian et~al.(2021)Bian, Liu, Meng, Xing, Xu, and Lu]{bian2021lipid}
Xueli Bian, Rui Liu, Ying Meng, Dongming Xing, Daqian Xu, and Zhimin Lu.
\newblock Lipid metabolism and cancer.
\newblock \emph{Journal of Experimental Medicine}, 218\penalty0 (1):\penalty0
  1--34, 2021.

\bibitem[Blei et~al.(2017)Blei, Kucukelbir, and McAuliffe]{Blei2017}
David~M. Blei, Alp Kucukelbir, and Jon~D. McAuliffe.
\newblock {Variational Inference: A Review for Statisticians}.
\newblock \emph{Journal of the American Statistical Association}, 112\penalty0
  (518):\penalty0 859--877, 2017.
\newblock \doi{10.1080/01621459.2017.1285773}.

\bibitem[Camerlenghi et~al.(2019)Camerlenghi, Dunson, Lijoi, Pr{\"{u}}nster,
  and Rodr{\'{i}}guez]{Camerlenghi2019}
Federico Camerlenghi, David~B. Dunson, Antonio Lijoi, Igor Pr{\"{u}}nster, and
  Abel Rodr{\'{i}}guez.
\newblock {Latent Nested Nonparametric Priors (with Discussion)}.
\newblock \emph{Bayesian Analysis}, 14\penalty0 (4):\penalty0 {1--54}, 2019.
\newblock \doi{10.1214/19-ba1169}.
\newblock URL \url{http://arxiv.org/abs/1801.05048}.

\bibitem[Capitoli et~al.(2025)Capitoli, van Abeelen, Piga, L’Imperio, Nobile,
  Besozzi, and Galimberti]{Capitoli2024}
Giulia Capitoli, Kirsten C.~J. van Abeelen, Isabella Piga, Vincenzo
  L’Imperio, Marco~S. Nobile, Daniela Besozzi, and Stefania Galimberti.
\newblock {Well Begun Is Half Done: The Impact of Pre-Processing in MALDI Mass
  Spectrometry Imaging Analysis Applied to a Case Study of Thyroid Nodules}.
\newblock \emph{Stats}, 8\penalty0 (3):\penalty0 1--14, 2025.

\bibitem[D'Angelo and Denti(2024)]{Dangelo2023}
Laura D'Angelo and Francesco Denti.
\newblock A finite-infinite shared atoms nested model for the bayesian analysis
  of large grouped data.
\newblock \emph{Bayesian Analysis}, In press:\penalty0 1--34, 2024.

\bibitem[Denti et~al.(2023)Denti, Camerlenghi, Guindani, and Mira]{Denti2021}
Francesco Denti, Federico Camerlenghi, Michele Guindani, and Antonietta Mira.
\newblock {A Common Atoms Model for the Bayesian Nonparametric Analysis of
  Nested Data}.
\newblock \emph{Journal of the American Statistical Association}, 118\penalty0
  (541):\penalty0 405--416, 2023.

\bibitem[Denti et~al.(2022)Denti, Capitoli, Piga, Clerici, Pagani, Criscuolo,
  Bindi, Principi, Chinello, Paglia, et~al.]{denti2022spatial}
Vanna Denti, Giulia Capitoli, Isabella Piga, Francesca Clerici, Lisa Pagani,
  Lucrezia Criscuolo, Greta Bindi, Lucrezia Principi, Clizia Chinello, Giuseppe
  Paglia, et~al.
\newblock Spatial multiomics of lipids, n-glycans, and tryptic peptides on a
  single ffpe tissue section.
\newblock \emph{Journal of Proteome Research}, 21\penalty0 (11):\penalty0
  2798--2809, 2022.

\bibitem[Denti et~al.(2024)Denti, Greco, Alviano, Capitoli, Monza, Smith,
  Pilla, Maggioni, Ivanova, Venetis, et~al.]{denti2024spatially}
Vanna Denti, Angela Greco, Antonio~Maria Alviano, Giulia Capitoli, Nicole
  Monza, Andrew Smith, Daniela Pilla, Alice Maggioni, Mariia Ivanova,
  Konstantinos Venetis, et~al.
\newblock Spatially resolved molecular characterization of noninvasive
  follicular thyroid neoplasms with papillary-like nuclear features (niftps)
  identifies a distinct proteomic signature associated with ras-mutant lesions.
\newblock \emph{International Journal of Molecular Sciences}, 25\penalty0
  (23):\penalty0 13115, 2024.

\bibitem[Durand et~al.(2022)Durand, Forbes, Phan, Truong, Nguyen, and
  Dama]{Durand2022}
J.~B. Durand, F.~Forbes, C.~D. Phan, L.~Truong, H.~D. Nguyen, and F.~Dama.
\newblock {Bayesian non-parametric spatial prior for traffic crash risk
  mapping: A case study of Victoria, Australia}.
\newblock \emph{Australian and New Zealand Journal of Statistics}, 64\penalty0
  (2):\penalty0 171--204, 2022.

\bibitem[Fu et~al.(2021)Fu, Zou, Shen, Nelson, Li, Wu, Yang, Zheng, Bruns,
  Zhao, Qin, and Dong]{Fu2021}
Yan Fu, Tiantian Zou, Xiaotian Shen, Peter~J. Nelson, Jiahui Li, Chao Wu,
  Jimeng Yang, Yan Zheng, Christiane Bruns, Yue Zhao, Lunxiu Qin, and Qiongzhu
  Dong.
\newblock Lipid metabolism in cancer progression and therapeutic strategies.
\newblock \emph{MedComm}, 2:\penalty0 27--59, 2021.
\newblock ISSN 2688-2663.
\newblock \doi{10.1002/mco2.27}.

\bibitem[Gobena et~al.(2024)Gobena, Admassu, Kinde, and
  Gessese]{gobena2024proteomics}
Semira Gobena, Bemrew Admassu, Mebrie~Zemene Kinde, and Abebe~Tesfaye Gessese.
\newblock Proteomics and its current application in biomedical area: Concise
  review.
\newblock \emph{The Scientific World Journal}, 2024\penalty0 (1):\penalty0
  4454744, 2024.

\bibitem[Govaert and Nadif(2013)]{govaert2013co}
G{\'e}rard Govaert and Mohamed Nadif.
\newblock \emph{Co-clustering: models, algorithms and applications}.
\newblock John Wiley \& Sons, 2013.

\bibitem[Guo et~al.(2019)Guo, Bemis, Rawlins, Agar, and
  Vitek]{guo2019unsupervised}
Dan Guo, Kylie Bemis, Catherine Rawlins, Jeffrey Agar, and Olga Vitek.
\newblock Unsupervised segmentation of mass spectrometric ion images
  characterizes morphology of tissues.
\newblock \emph{Bioinformatics}, 35\penalty0 (14):\penalty0 i208--i217, 2019.

\bibitem[Hsieh et~al.(2017)Hsieh, Purdue, Signoretti, Swanton, Albiges,
  Schmidinger, Heng, Larkin, and Ficarra]{hsieh2017renal}
James~J Hsieh, Mark~P Purdue, Sabina Signoretti, Charles Swanton, Laurence
  Albiges, Manuela Schmidinger, Daniel~Y Heng, James Larkin, and Vincenzo
  Ficarra.
\newblock Renal cell carcinoma.
\newblock \emph{Nature reviews Disease primers}, 3\penalty0 (1):\penalty0
  1--19, 2017.

\bibitem[Hsieh et~al.(2018)Hsieh, Le, Cao, Cheng, and
  Creighton]{hsieh2018genomic}
James~J Hsieh, Valerie Le, Dengfeng Cao, Emily~H Cheng, and Chad~J Creighton.
\newblock Genomic classifications of renal cell carcinoma: A critical step
  towards the future application of personalized kidney cancer care with
  pan-omics precision.
\newblock \emph{The Journal of pathology}, 244\penalty0 (5):\penalty0 525--537,
  2018.

\bibitem[Krestensen et~al.(2023)Krestensen, Heeren, and
  Balluff]{krestensen2023state}
Kasper~K Krestensen, Ron~MA Heeren, and Benjamin Balluff.
\newblock State-of-the-art mass spectrometry imaging applications in biomedical
  research.
\newblock \emph{Analyst}, 148\penalty0 (24):\penalty0 6161--6187, 2023.

\bibitem[Lee et~al.(2013)Lee, M{\"{u}}ller, Zhu, and Ji]{Lee2013}
Juhee Lee, Peter M{\"{u}}ller, Yitan Zhu, and Yuan Ji.
\newblock {A nonparametric Bayesian model for local clustering with application
  to proteomics}.
\newblock \emph{Journal of the American Statistical Association}, 108\penalty0
  (503):\penalty0 775--788, 2013.
\newblock ISSN 01621459.
\newblock \doi{10.1080/01621459.2013.784705}.

\bibitem[Lijoi et~al.(2023)Lijoi, Pr{\"{u}}nster, and Rebaudo]{lijoi2023}
Antonio Lijoi, Igor Pr{\"{u}}nster, and Giovanni Rebaudo.
\newblock {Flexible clustering via hidden hierarchical Dirichlet priors}.
\newblock \emph{Scandinavian Journal of Statistics}, 50\penalty0 (1):\penalty0
  213--234, 2023.
\newblock ISSN 14679469.
\newblock \doi{10.1111/sjos.12578}.

\bibitem[Linderman et~al.(2015)Linderman, Johnson, and
  Adams]{linderman2015multinomial}
Scott~W. Linderman, Matthew~J. Johnson, and Ryan~P. Adams.
\newblock Dependent multinomial models made easy: Stick‑breaking with the
  pólya‑gamma augmentation.
\newblock In \emph{Advances in Neural Information Processing Systems 28}, pages
  3456--3464, 2015.

\bibitem[L{\"{u}} et~al.(2020)L{\"{u}}, Arbel, and Forbes]{Lu2020}
Hongliang L{\"{u}}, Julyan Arbel, and Florence Forbes.
\newblock {Bayesian nonparametric priors for hidden Markov random fields}.
\newblock \emph{Statistics and Computing}, 30\penalty0 (4):\penalty0
  1015--1035, 2020.

\bibitem[Maechler et~al.(2025)Maechler, Rousseeuw, Struyf, Hubert, and
  Hornik]{cluster_package}
Martin Maechler, Peter Rousseeuw, Anja Struyf, Mia Hubert, and Kurt Hornik.
\newblock \emph{cluster: Cluster Analysis Basics and Extensions}, 2025.

\bibitem[Malsiner-Walli et~al.(2016)Malsiner-Walli, Fr{\"u}hwirth-Schnatter,
  and Gr{\"u}n]{MalsinerWalli2016}
Gertraud Malsiner-Walli, Sylvia Fr{\"u}hwirth-Schnatter, and Bettina Gr{\"u}n.
\newblock {Model-based Clustering Based on Sparse Finite Gaussian Mixtures}.
\newblock \emph{Statistics and Computing}, 26:\penalty0 303--324, 2016.
\newblock \doi{10.1007/s11222-014-9500-2}.

\bibitem[McGrory et~al.(2009)McGrory, Titterington, Reeves, and
  Pettitt]{McGrory2009}
C.~A. McGrory, D.~M. Titterington, R.~Reeves, and A.~N. Pettitt.
\newblock {Variational bayes for estimating the parameters of a hidden potts
  model}.
\newblock \emph{Statistics and Computing}, 19\penalty0 (3):\penalty0 329--340,
  2009.

\bibitem[Moore et~al.(2023)Moore, Patterson, Norris, and
  Caprioli]{moore2023prospective}
Jessica~L Moore, Nathan~Heath Patterson, Jeremy~L Norris, and Richard~M
  Caprioli.
\newblock Prospective on imaging mass spectrometry in clinical diagnostics.
\newblock \emph{Molecular \& Cellular Proteomics}, 22\penalty0 (9):\penalty0
  1--10, 2023.

\bibitem[Moores et~al.(2020)Moores, Nicholls, Pettitt, and
  Mengersen]{Moores2020}
Matthew Moores, Geoff~K. Nicholls, Anthony~N. Pettitt, and Kerrie Mengersen.
\newblock {Scalable bayesian inference for the inverse temperature of a hidden
  potts model}.
\newblock \emph{Bayesian Analysis}, 15\penalty0 (1):\penalty0 1--27, 2020.

\bibitem[Murray et~al.(2006)Murray, Ghahramani, and MacKay]{Murray2006}
Iain Murray, Zoubin Ghahramani, and David~J.C. MacKay.
\newblock {MCMC for doubly-intractable distributions}.
\newblock \emph{Proceedings of the 22nd Conference on Uncertainty in Artificial
  Intelligence, UAI 2006}, pages 359--366, 2006.

\bibitem[Prasad et~al.(2022)Prasad, Postma, Franceschi, Buydens, and
  Jansen]{prasad2022evaluation}
Mridula Prasad, Geert Postma, Pietro Franceschi, Lutgarde~MC Buydens, and
  Jeroen~J Jansen.
\newblock Evaluation and comparison of unsupervised methods for the extraction
  of spatial patterns from mass spectrometry imaging data (msi).
\newblock \emph{Scientific Reports}, 12\penalty0 (1):\penalty0 15687, 2022.

\bibitem[Rebaudo et~al.(2021)Rebaudo, Lin, and Mueller]{Lin2021}
Giovanni Rebaudo, Qiaohui Lin, and Peter Mueller.
\newblock {Separate Exchangeability as Modeling Principle in Bayesian
  Nonparametrics}.
\newblock \emph{ArXiv Preprint}, 2021.

\bibitem[Rodr{\'{i}}guez et~al.(2008)Rodr{\'{i}}guez, Dunson, and
  Gelfand]{Rodriguez2008}
Abel Rodr{\'{i}}guez, David~B. Dunson, and Alan~E. Gelfand.
\newblock {The nested dirichlet process}.
\newblock \emph{Journal of the American Statistical Association}, 103\penalty0
  (483):\penalty0 1131--1154, 2008.
\newblock ISSN 01621459.
\newblock \doi{10.1198/016214508000000553}.

\bibitem[Rohner et~al.(2005)Rohner, Staab, and Stoeckli]{rohner2005maldi}
T.C. Rohner, D.~Staab, and M.~Stoeckli.
\newblock Maldi mass spectrometric imaging of biological tissue sections.
\newblock \emph{Mechanisms of Ageing and Development}, 126\penalty0
  (1):\penalty0 177--185, 2005.

\bibitem[Shafiee et~al.(2020)Shafiee, Ortori, Barrett, Mongan, Abu, and
  Atiomo]{Shafiee2020}
Mohamad~Nasir Shafiee, Catharine~A. Ortori, David~A. Barrett, Nigel~P. Mongan,
  Jafaru Abu, and William Atiomo.
\newblock Lipidomic biomarkers in polycystic ovary syndrome and endometrial
  cancer.
\newblock \emph{International Journal of Molecular Sciences}, 21:\penalty0
  1--15, 2020.
\newblock ISSN 14220067.
\newblock \doi{10.3390/ijms21134753}.

\bibitem[Stanley et~al.(2022)Stanley, Moremen, Lewis, Taniguchi, and
  Aebi]{stanley2022n}
Pamela Stanley, Kelley~W. Moremen, Nathan~E. Lewis, Naoyuki Taniguchi, and
  Markus Aebi.
\newblock \emph{N-Glycans}.
\newblock Cold Spring Harbor Laboratory Press, Cold Spring Harbor (NY), 4th
  edition, 2022.
\newblock ISBN 9781621824220.
\newblock URL \url{http://europepmc.org/books/NBK579964}.

\bibitem[Tan and Witten(2014)]{tan2014sparse}
Kean~Ming Tan and Daniela~M Witten.
\newblock Sparse biclustering of transposable data.
\newblock \emph{Journal of Computational and Graphical Statistics}, 23\penalty0
  (4):\penalty0 985--1008, 2014.

\bibitem[Tibshirani et~al.(2001)Tibshirani, Walther, and
  Hastie]{tibshirani2001estimating}
Robert Tibshirani, Guenther Walther, and Trevor Hastie.
\newblock Estimating the number of clusters in a data set via the gap
  statistic.
\newblock \emph{{Journal of the Royal Statistical Society: Series B}},
  63\penalty0 (2):\penalty0 411--423, 2001.

\bibitem[Tuck et~al.(2022)Tuck, Grélard, Blanc, and Desbenoit]{Tuck2022}
Michael Tuck, Florent Grélard, Landry Blanc, and Nicolas Desbenoit.
\newblock Maldi-msi towards multimodal imaging: Challenges and perspectives.
\newblock \emph{Frontiers in Chemistry}, 10:\penalty0 1--11, 2022.
\newblock ISSN 2296-2646.
\newblock \doi{10.3389/fchem.2022.904688}.
\newblock URL \url{http://dx.doi.org/10.3389/fchem.2022.904688}.

\bibitem[Vasseur and Guillaumond(2022)]{vasseur2022lipids}
Sophie Vasseur and Fabienne Guillaumond.
\newblock Lipids in cancer: a global view of the contribution of lipid pathways
  to metastatic formation and treatment resistance.
\newblock \emph{Oncogenesis}, 11\penalty0 (1):\penalty0 46, 2022.

\bibitem[Verbeeck et~al.(2019)Verbeeck, Caprioli, and Van~de
  Plas]{Verbeeck2019}
Nico Verbeeck, Richard~M. Caprioli, and Raf Van~de Plas.
\newblock Unsupervised machine learning for exploratory data analysis in
  imaging mass spectrometry.
\newblock \emph{Mass Spectrometry Reviews}, 39\penalty0 (3):\penalty0
  245–291, October 2019.
\newblock ISSN 1098-2787.
\newblock \doi{10.1002/mas.21602}.
\newblock URL \url{http://dx.doi.org/10.1002/mas.21602}.

\bibitem[Wallace et~al.(2024)Wallace, West, McDowell, Lu, Bruner, Mehta,
  Aoki-Kinoshita, Angel, and Drake]{wallace2024n}
Elizabeth~N Wallace, Connor~A West, Colin~T McDowell, Xiaowei Lu, Evelyn
  Bruner, Anand~S Mehta, Kiyoko~F Aoki-Kinoshita, Peggi~M Angel, and Richard~R
  Drake.
\newblock An n-glycome tissue atlas of 15 human normal and cancer tissue types
  determined by maldi-imaging mass spectrometry.
\newblock \emph{Scientific Reports}, 14\penalty0 (1):\penalty0 489, 2024.

\bibitem[Zhang et~al.(2024)Zhang, Lu, Ebbini, Huang, Lu, and Li]{zhang2024mass}
Hua Zhang, Kelly~H Lu, Malik Ebbini, Penghsuan Huang, Haiyan Lu, and Lingjun
  Li.
\newblock Mass spectrometry imaging for spatially resolved multi-omics
  molecular mapping.
\newblock \emph{npj Imaging}, 2\penalty0 (1):\penalty0 20, 2024.

\bibitem[Zhvansky et~al.(2021)Zhvansky, Sorokin, Shurkhay, Zavorotnyuk,
  Bormotov, Pekov, Potapov, Nikolaev, and Popov]{zhvansky2021comparison}
Evgeny Zhvansky, Anatoly Sorokin, Vsevolod Shurkhay, Denis Zavorotnyuk, Denis
  Bormotov, Stanislav Pekov, Alexander Potapov, Evgeny Nikolaev, and Igor
  Popov.
\newblock Comparison of dimensionality reduction methods in mass spectra of
  astrocytoma and glioblastoma tissues.
\newblock \emph{Mass Spectrometry}, 10\penalty0 (1):\penalty0 A0094, 2021.

\end{thebibliography}
